\documentclass[apj,twocolumn,twocolappendix,numberedappendix]{openjournal}
\usepackage{newtxtext,newtxmath,enumitem,multirow,ctable,verbatim}
\usepackage[T1]{fontenc}
\usepackage[breaklinks,colorlinks,citecolor=blue,urlcolor=blue]{hyperref}

\newcommand{\Alf}{{Alfv\'en}}

\newcommand{\orcidauthor}[3]{\author{\href{http://orcid.org/#1}{#2$^{#3}$}}}
\usepackage{pifont}% http://ctan.org/pkg/pifont
\newcommand{\cmark}{\ding{51}}%
\newcommand{\xmark}{\ding{55}}%

\shorttitle{Masers Favor Magnetically-Dominated Disks}
\shortauthors{Hopkins, Baron, \&\ Piotrowska}

\begin{document}
\title{\vspace{-0.8cm}Masers and Broad-Line Mapping Favor Magnetically-Dominated AGN Accretion Disks\vspace{-1.5cm}}
\orcidauthor{0000-0003-3729-1684}{Philip F. Hopkins}{1,*}
\orcidauthor{0000-0003-4974-3481}{Dalya Baron}{2}, 
\orcidauthor{0000-0003-1661-2338}{Joanna M. Piotrowska}{1}
\affiliation{$^{1}$TAPIR, Mailcode 350-17, California Institute of Technology, Pasadena, CA 91125, USA}
\affiliation{$^{2}$Kavli Institute for Particle Astrophysics \& Cosmology, Stanford University, CA 94305, USA}
\thanks{$^*$E-mail: \href{mailto:phopkins@caltech.edu}{phopkins@caltech.edu}},

\begin{abstract}
We present a novel and powerful constraint on the physics of supermassive black hole (BH) accretion disks. We show that in the outer disk (radii $R \gtrsim 0.01\,$pc or $\gtrsim 1000\,R_{G}$), models supported by thermal or radiation pressure predict disk masses which are much larger than the BH mass and increase with radius, yielding rapidly-rising, extremely non-Keplerian rotation curves. More generally, we show that any observational upper limit to the deviation from Keplerian potentials at these radii directly constrains the physical form of the pressure in disks. We then show that existing maser and broad line region (BLR) kinematic observations immediately rule out the classic thermal-pressure-dominated Shakura Sunyaev-like $\alpha$-disk model, and indeed rule out any thermal or radiation (or cosmic-ray) pressure-dominated disk, as the required temperatures and luminosities of the gas at large radii would exceed those observed by orders of magnitude. We show that models where the pressure comes entirely from turbulence (without thermal, radiation, or magnetic sources) could in principle be viable but would require turbulent Toomre $Q \gtrsim 100$, far larger than predicted by self gravitating/gravito-turbulent models. However, recently proposed models of magnetic pressure-dominated disks agree with all of the observational constraints. These magnetically-dominated models also appear to agree better with constraints on maser magnetic fields, compared to the other possibilities. Observations appear to strongly favor the hypothesis that the outer regions of BH accretion disks are in the ``hyper-magnetized'' state.
\end{abstract}
\maketitle

\section{Introduction}
\label{sec:intro}

Quasars and active galactic nuclei (AGN) are powered by accretion disks around supermassive BHs \citep{schmidt:1963.qso.redshift,soltan82}. The overwhelming majority of accretion disk models assume that the accretion disk itself is thermal and/or radiation-pressure dominated, and so is qualitatively similar to the classic \citet[SS73]{shakurasunyaev73} and \citet{novikov.thorne:1973.astro.of.bhs}-like $\alpha$-disk model. This includes many variant disk models including thin or ``slim,'' magnetically elevated, and advection-dominated \citep{frank:2002.accretion.book,abramowicz:accretion.theory.review}. Recently, however, simulations following magnetized gas inflows self-consistently from star-forming ISM through accretion-disk scales in  \citet{hopkins:superzoom.overview,hopkins:superzoom.disk,hopkins:superzoom.imf} found a qualitatively different type of accretion disk emerges: so-called ``hyper-magnetized'' or ``flux-frozen'' disks, whose midplane pressure is dominated by toroidal magnetic fields amplified and stretched from ISM fields. This has now been seen in other simulation contexts including lower-rate accretion onto black holes in elliptical galaxies \citep{guo:2024.fluxfrozen.disks.lowmdot.ellipticals}; 
intermediate-mass to $\sim 10^{6}\,{\rm M}_{\odot}$ BHs in dense star clusters resembling ``little red dots''  \citep{shi:2024.imbh.growth.feedback.survey,shi:2024.seed.to.smbh.case.study.subcluster.merging.pairing.fluxfrozen.disk}; 
some magnetized first-stars simulations \citep{luo:2024.magnetically.dominated.disk.like.our.zoomins.zoomin.on.first.supermassive.star.situation}; 
circum-binary magnetized cloud collapse \citep{wang:2025.hypermagnetized.circumbinary.disk.flux.frozen.cavity.to.pc.scales}; 
followups to smaller (near-horizon) radii \citep{kaaz:2024.hamr.forged.fire.zoom.to.grmhd.magnetized.disks,hopkins:superzoom.agn.disks.to.isco.with.gizmo.rad.thermochemical.properties.nlte.multiphase.resolution.studies}; 
as well as older simulations of more idealized setups \citep{gaburov:2012.public.moving.mesh.code} and similar recent studies \citep{squire:2024.mri.shearing.box.strongly.magnetized.different.beta.states,guo:2025.idealized.sphere.collapse.sims.hypermagnetized.disks.resolution.dependent.on.resolving.thermal.scale.height.but.limited.physics}. 
Analytic models \citep{hopkins:superzoom.analytic,hopkins:multiphase.mag.dom.disks} have argued that such disks should be ubiquitous around high-accretion rate SMBHs, and have properties unlike traditional $\alpha$-disks. 

One of the most dramatic differences between these hyper-magnetized/flux-frozen (plasma $\beta \equiv P_{\rm thermal}/P_{\rm magnetic} \ll 1$) and traditional ($\beta \gg 1$) $\alpha$-disks is in the disk mass, especially at large radii/distances from the BH. Because the Maxwell stresses are, by definition, strong in the hyper-magnetized disks, accretion is fundamentally dynamical, i.e.\ the inflow timescale ($t_{\rm acc} \sim M_{\rm disk}/\dot{M}$) at distances from the BH radius of influence (BHROI) to horizon scales is of order the orbital time ($t_{\rm acc} \sim 10\,\Omega^{-1}$ in terms of the disk orbital frequency $\Omega$). In contrast, in an $\alpha$-disk, $t_{\rm acc} \sim \alpha^{-1}\,(V_{\rm c}/c_{s})^{2}\,\Omega^{-1} \sim 10^{6}-10^{8}\, \Omega^{-1}$ in the outer disk ($\sim 0.01-10\,$pc), i.e.\ accretion proceeds secularly over millions of orbits at large radii. A necessary consequence of this is that, in order to provide a given observed accretion rate and/or luminosity, the outer disk must be orders-of-magnitude more massive in $\alpha$-disk models compared to hyper-magnetized disks. At radii $\gtrsim 0.01$\,pc, this in turn leads to a major qualitative difference: $\alpha$ disks (whether thermal-or-radiation pressure dominated) are predicted to be much more massive than the BH itself ($M_{\rm disk}(<R) \sim \pi \Sigma_{\rm gas} R^{2} \gg M_{\rm BH}$), meaning that the gravitational potential becomes qualitatively modified and the rotation curve or circular velocity $V_{\rm c} \equiv \sqrt{G\,M_{\rm total}(<R)/R}$ goes from the standard declining Keplerian behavior ($V_{\rm c} \propto R^{-1/2}$) to \textit{rising} ($V_{\rm c} \propto R^{+3/8}$ or $R^{+2/3}$, like in the central few kpc of galaxies). In contrast, hyper-magnetized disks have $M_{\rm disk}(<R) \ll M_{\rm BH}$ so the potential remains approximately Keplerian out to the BHROI $R_{\rm BHROI} \equiv G\,M_{\rm BH}/\sigma_{\rm gal}^{2} \sim 10\,{\rm pc}\,(M_{\rm BH}/10^{8}\,{\rm M_{\odot}})\,(\sigma_{\rm gal}/200\,{\rm km\,s^{-1}})^{-2}$ (in terms of the galactic velocity dispersion $\sigma_{\rm gal}$) exterior to which, by definition, the galactic potential of stars and dark matter dominate over the BH. 

This range of radii falls well within the range where excellent constraints on the kinematics and rotation curves of gas around many AGN exist, from a combination of maser, infrared (dust) and optical (broad-line) interferometric, and reverberation-mapping constraints mapping out the dynamics of gas as a function of BH-centric radius $R$. In this manuscript, we show that this leads to robust qualitative constraints on the form of the pressure supporting the accretion disk, and the existing observations very clearly rule out thermal or radiation pressure-dominated accretion disks at $R \gtrsim 0.01\,$pc around most accreting SMBHs. We show that, completely independent of detailed assumptions of the accretion disk model, \textit{any} thermal or radiation or cosmic-ray pressure dominated accretion disk at these radii compatible with the observations of kinematics would grossly violate other fundamental observational constraints (e.g.\ predicting orders-of-magnitude larger luminosities than observed). We also show that many of these models appear to be in tension with upper limits on the magnetic field strengths observed at these radii. In contrast, we show that the existing predictions of simple analytic similarity models for flux-frozen, hyper-magnetized disks appear to agree naturally with all of these observational constraints.

\begin{figure*}
	\centering
	\includegraphics[width=0.49\textwidth]{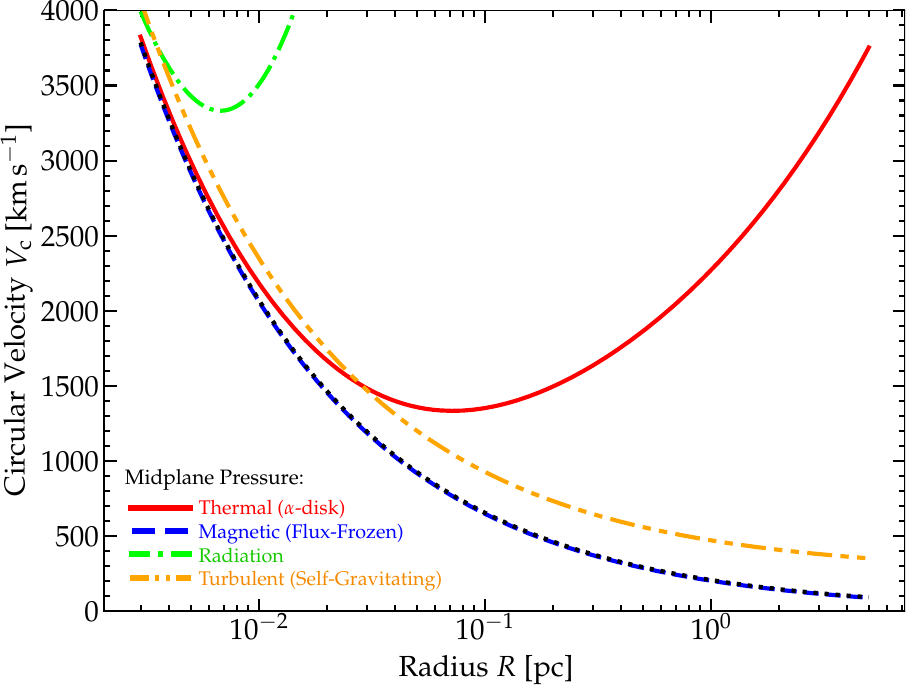} \hspace{3pt}
	\includegraphics[width=0.48\textwidth]{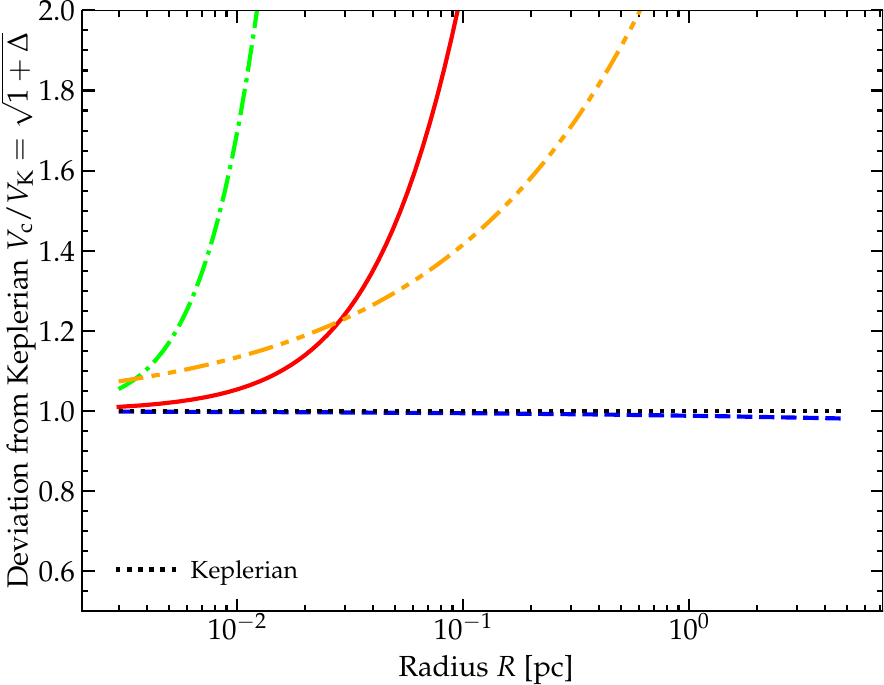} 
	\caption{Expected scaling of gas circular velocity with radius and its deviation from Keplerian motion for different accretion disk models. 
	Lines correspond to standard accretion disk models assuming a $M_{\rm BH}=10^{7}\,{\rm M_{\odot}}$ accreting near-Eddington ($m_{7}=\dot{m}=1$) with an effective accretion-stress-to-total-pressure ratio $\alpha=0.1$. 
	We assume the disk total pressure is primarily: (1) thermal (a standard SS73-like or $\alpha$-disk; \S~\ref{sec:thermal}); (2) magnetic (a flux-frozen or hyper-magnetized disk; \S~\ref{sec:magnetic}); (3) radiation (a slim or radiation-pressure-supported disk; \S~\ref{sec:radiation}); (4) turbulent (constant-$Q_{0}$ or gravito-turbulent, with $Q_{0}=0.04$ as required to fit some NGC 1068 observations with this model; \S~\ref{sec:turb}). 
	For each, we plot the predicted $V_{c}^{2}(R) \equiv G M_{\rm enc}(<R)/R = G\,[M_{\rm BH} + M_{\rm disk}(<R)]/R$. 
	We compare to Keplerian (in absolute units at \textit{left} or relative \textit{right}). 
	Per \S~\ref{sec:theory}, different assumptions about what dominates the pressure in the outer disk, for a given BH mass and luminosity, give very different predictions for the disk mass, with larger disk masses producing large deviations from Keplerian circular velocity curves (even rapidly-rising curves). 
	Attempting to ``re-tune'' the thermal or radiation models by arbitrarily changing the predicted temperatures or opacities to suppress the deviations from Keplerian lead to other, immediately ruled-out predictions (like $\sim 10$ orders of magnitude larger luminosities). 
	\label{fig:vc}}
\end{figure*}

\section{Theoretical Basis}
\label{sec:theory}

\begin{figure*}
	\centering
	\includegraphics[width=0.48\textwidth]{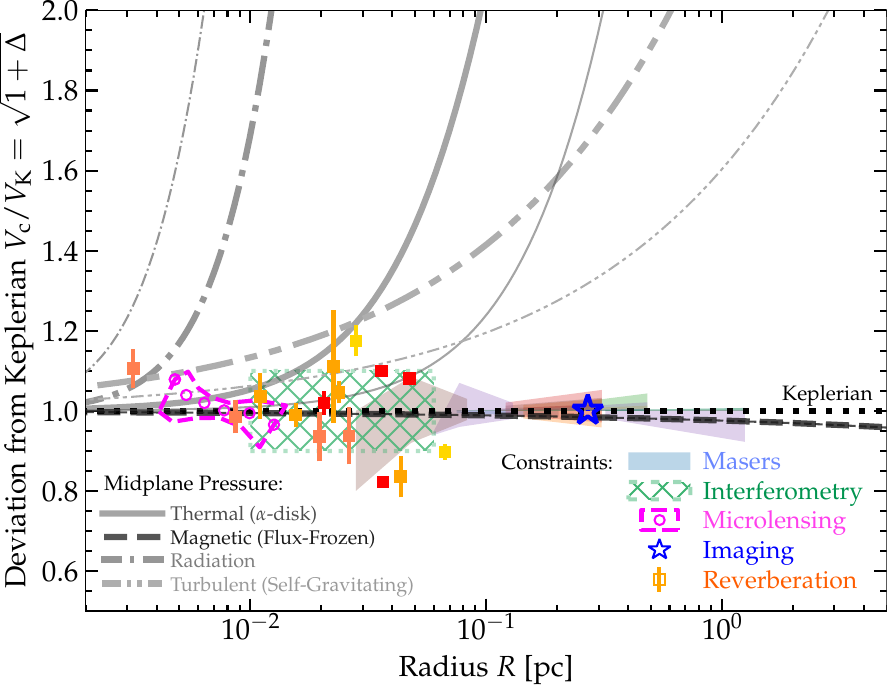} 
	\includegraphics[width=0.49\textwidth]{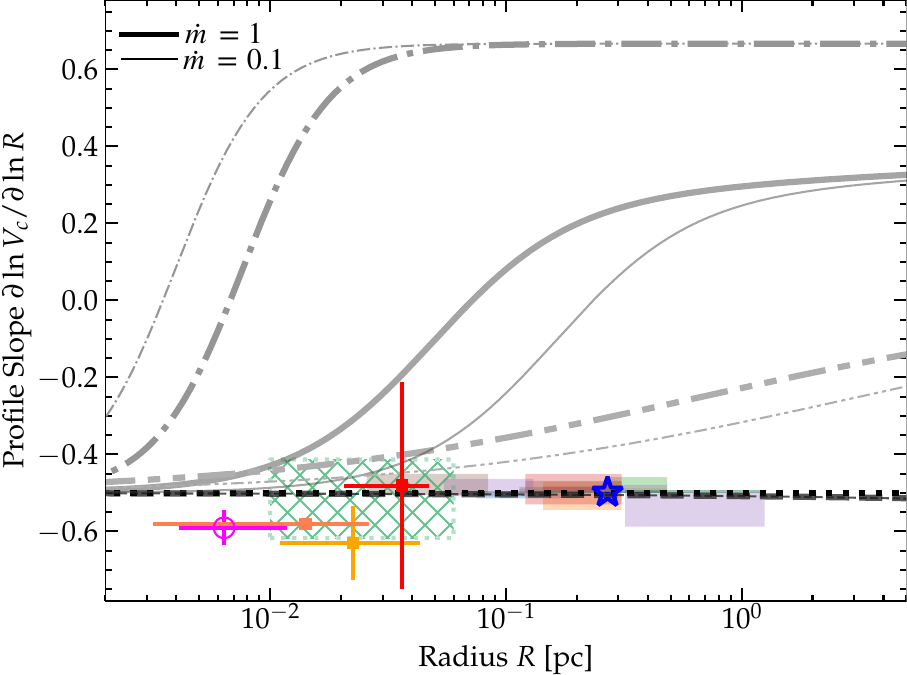} 
	\caption{Observations (\textit{color}; \S~\ref{sec:obs}), compared to the models (\textit{grayscale}) per Fig.~\ref{fig:vc}. 
	We compile constraints from kinematics/rotation-curve-fitting from compilations of masers (\textit{shaded}); GRAVITY optical interferometry of the BLR (\textit{hatched}); microlensing response in broad-line wings (\textit{circles}); plus individual resolved neutral-gas imaging and gas mass measurements in Circinus (\textit{star}) and coordinated reverberation mapping of different BLR lines (\textit{squares}). 
	We compare model predictions assuming different midplane pressure sources as Fig.~\ref{fig:vc}, for $\alpha=0.1$, $m_{7}=1$ and $\dot{m}=1$ (\textit{thick}) or $\dot{m}=0.1$ (\textit{thin}). 
	\textit{Left:} Deviation from Keplerian circular velocities. 
	\textit{Right:} Logarithmic slope of the circular velocity curve, fitted to each over the range of measured velocities. 
	The best masers are within $\sim 1\%$ of Keplerian, while the best BLR measurements constrain the profile to within $\sim 5-10\%$ of Keplerian, sufficient to rule out radiation or thermal-pressure dominated models, or turbulence-dominated models with $Q \lesssim 1$ at $r \gtrsim 0.003\,$pc, but consistent with magnetic models at these radii. There is a hint of steeper-than-Keplerian curves in some BLR and maser systems, which only occurs in the magnetic models. 
	\label{fig:vc.obs}}
\end{figure*}

As emphasized in \S~\ref{sec:intro}, different families of accretion disk models result in different scalings of circular velocity with distance from the black hole (shown in Fig.~\ref{fig:vc}). 
In what follows, we will consider different models for the accretion disk pressure (thermal, radiation, magnetic, turbulent, cosmic-ray) in turn. For each, we will present the ``standard'' literature model, using the specific scalings from canonical papers deriving models under the assumption that the given pressure is dominant in the midplane (summarized in Figs.~\ref{fig:vc}-\ref{fig:B}). But we will then consider much more general constraints, asking ``what if'' we removed normal self-consistency assumptions and simply allowed that pressure (e.g.\ the thermal pressure or temperature) to have any value we wanted, in order to show that most models cannot possibly fit kinematic observations without severely violating other basic observational constraints. It is therefore helpful to define some basic background terms and scalings.

The defining assumption generic to accretion disk models is that some stress or torque removes angular momentum at a rate balancing the change in orbital angular momentum in the disk \citep[e.g.][]{abramowicz:accretion.theory.review}. This can be written as $\dot{M} = 3\pi\nu_{\rm v,\,eff}\Sigma_{\rm gas}$ in terms of the accretion rate $\dot{M}$, disk surface density $\Sigma_{\rm gas}$, and effective viscosity $\nu_{\rm v,\,eff}\equiv \varpi_{\nu}\,V_{e}^{2}/\Omega \equiv \Pi /\Omega\rho$ where $\rho$ is the midplane gas density, $\Omega \equiv V_{\rm c}/R$ at some distance $R$ in a potential with circular velocity $V_{\rm c}^{2} \equiv G M_{\rm enc}(<R)/R$ (total enclosed mass $M_{\rm enc}(<R)$ -- note we do not assume a Keplerian potential) and either some effective transport (turbulent, \Alf, or other) velocities $V_{e}$ or relevant component of the stress tensor $\Pi$ driving accretion, with $\varpi_{\nu}$ some order-unity prefactor that depends on details of e.g.\ the vertical disk profile.\footnote{For our purposes these factors $\varpi$ vary quite weakly -- by tens of percents or so -- compared to the orders-of-magnitude differences between different models for the disk pressure, so we can safely treat them as unity here.} Since the total disk mass enclosed is an increasing function of $R$, we can integrate over $2\pi R dR$ from $R=0$ to $R$, and obtain:
\begin{align}
M_{\rm gas}(<R) \approx 3 \varpi \,\left( \frac{\Pi_{\rm eff}}{\rho} \right)^{-1}\dot{M} \Omega R^{2}\ .
\end{align}
\medskip
We emphasize that this result assumes a steady-state with $\dot{M}$ constant as a function of $R$. This is the standard assumption in the accretion disk literature on the scales of interest, and is appropriate for disks in statistical equilibrium fed at rate $\dot{M}$ from large radii and accreting at the same $\dot{M}$ onto the horizon. One might ask whether our constraints can be evaded by models in which $\dot{M}(R)$ varies strongly with radius -- for example, models in which the large-radius $\dot{M}(R) > \dot{M}_{\rm horizon}$, with mass lost to outflows or star formation before reaching the horizon (as in radiation-driven-wind scenarios for super-Eddington disks, or in fragmentation-ablation scenarios; see \S~\ref{sec:radiation}-\ref{sec:pressures}). Such scenarios actually make our constraints {\em stronger}, not weaker: the steady-state disk mass at radius $R$ needed to supply the {\em observed} horizon-level accretion rate $\dot{M}_{\rm horizon}$ is enhanced by a factor $\dot{M}(R)/\dot{M}_{\rm horizon}$, so the required $\Pi_{\rm eff}/\rho$ and all derived limits (on $T$, $L$, $n$) become correspondingly more extreme.

Again, this is true by definition for any accretion disk model. 
Now, combining this with the definition of $V_{\rm c}$ and the Keplerian velocity $V_{K}$ ($V_{\rm c}$ if $M_{\rm enclosed}(<R)$ were equal to $M_{\rm BH}$): 
\begin{align}
\label{eqn:vc} V_{c} &\equiv \sqrt{\frac{G M_{\rm enclosed}(<R)}{R}} \ , \\
V_{K} &\equiv \sqrt{\frac{G M_{\rm BH}}{R}} \ , 
\end{align}
As a technical point, Eq.~\ref{eqn:vc} is the expression for the circular velocity in a spherically-symmetric mass distribution; for a thin disk it is strictly correct only for a Mestel surface-density profile \citep[see e.g.][\S\,2.6]{binneytremaine}. For the actual disk surface-density profiles considered here (SS73-like $\Sigma \propto R^{-3/4}$ or $R^{-1}$, radiation-dominated, or Kuzmin), the correction to $V_{c}$ for the disk-vs-spherical potential is at most $\lesssim 10\%$ even in the limiting infinitely-flattened Kuzmin case \citep{binneytremaine}, and has the sign that the actual disk $V_{c}$ is {\em larger} than Eq.~\ref{eqn:vc} for the same $M(<R)$, because more mass is concentrated in the midplane at any given radius. The correction is thus well below the uncertainties of the models plotted in Figs.~\ref{fig:vc}-\ref{fig:vc.obs}, and moreover acts to strengthen -- rather than weaken -- our rule-out of thermal and radiation-pressure-dominated disks. We therefore retain Eq.~\ref{eqn:vc} throughout, with the understanding that the corresponding corrections for a disk potential only make our conclusions more robust.

we have
\begin{align}
\frac{V_{c}^{2}}{V_{K}^{2}} &\approx \frac{M_{\rm BH} + M_{\rm gas}(<R)}{M_{\rm BH}} 
\approx 1 + \psi \left[\sqrt{1+\frac{\psi^{2}}{4}} + \frac{\psi}{2} \right] \ , 
\end{align}
where the dimensionless quantity $\psi$ is given by 
\begin{align}
\nonumber \psi &\equiv 3 \varpi \,\left( \frac{\Pi_{\rm eff}}{\rho} \right)^{-1} \frac{\dot{M} V_{K} R}{M_{\rm BH}} \sim \frac{3 L}{\epsilon_{r} c^{2}} \left( \frac{\Pi_{\rm eff}}{\rho} \right)^{-1} \sqrt{\frac{G R}{M_{\rm BH}}} \\
&\sim \frac{3 \dot{m}}{t_{S}} \left( \frac{\Pi_{\rm eff}}{\rho} \right)^{-1} \sqrt{G M_{\rm BH} R} \\ 
\nonumber &\sim 4\,\left( \frac{\Pi_{\rm eff}}{\rho \, ({\rm km\,s^{-1}})^{2}}  \right)^{-1}\,\dot{m} m_{7}^{1/2} r_{0.1}^{1/2} \ .
\end{align}
In the last equations we define some convenient units: $L=\epsilon_{r} \dot{M} c^{2}$, Salpeter time $t_{S} \approx 5\times10^{7}\,{\rm yr}$, $r_{0.1} \equiv R/0.1\,{\rm pc}$, $m_{7} \equiv M_{\rm BH}/10^{7}\,{\rm M_{\odot}}$, and $\dot{m} \equiv 0.1\,\dot{M}\,c^{2}/L_{\rm Edd}$ (i.e.\ accretion rate relative to Eddington for a radiative efficiency of $\epsilon_{r} \sim 0.1$). 

Let us define the upper limit on the deviation of the circular velocity from Keplerian as:
\begin{align}
V_{c}^{2} \le V_{K}^{2}\,\left( 1 + \Delta \right)
\end{align}
i.e.\ $(V_{c}^{2}-V_{K}^{2})/V_{K}^{2} \le \Delta$ (or equivalently if $V_{c}/V_{K} < 1+\delta$, $\Delta = 2\,\delta + \delta^{2}$). 
This is equivalent to $\psi \le \Delta/\sqrt{1+\Delta}$, or 
\begin{align}
\label{eqn:constraint} \frac{P_{\rm tot}}{\rho} &> \frac{\Pi_{\rm eff}}{\rho} >  \frac{3\dot{m} \sqrt{G M_{\rm BH} R (1+\Delta)}} {t_{S} \Delta} \ .
\end{align}
In other words, any observed upper limit to $\Delta$, i.e.\ close-to-Keplerian behavior of the rotation curve, sets a \textit{lower} limit to the ratio $P_{\rm tot}/\rho$, i.e.\ the specific pressure-per-unit-mass, needed to explain the observed accretion rate or luminosity in \textit{any} self-consistent accretion disk model. Conversely, a given accretion disk model, which assumes some source of pressure $\Pi/\rho$ in the midplane, predicts a \textit{lower} limit to $\Delta$, i.e.\ a minimum mass of the accretion disk needed to power the observed accretion, and therefore a minimum deviation from Keplerian rotation (Fig.~\ref{fig:vc}).

In the last equation, we have used the fact that in any reasonable model, $\Pi_{\rm eff} < P_{\rm tot}$ -- i.e.\ the salient net stress component causing dissipation and angular momentum transfer cannot significantly exceed the total stress/pressure. At any given radius, we can simply define 
\begin{align}
\Pi_{\rm eff} \equiv \alpha P_{\rm tot} \lesssim P_{\rm tot} \ , 
\end{align}
(i.e.\  $\alpha \lesssim 1$). Note we define this parameter by analogy to the classic $\alpha$-disk model, but those models specifically assume $\alpha$ is a constant (we allow it to be a function of radius or other properties) and that the pressure is primarily thermal. So we stress in our case that this is a convenient parameterization, with no loss of generality.

Note that because $\Delta \ll 1$ for the cases of greatest interest, i.e.\ the potentials are not far from Keplerian, we can also use the above scalings with the fact that the scale height of the disk in a Keplerian potential $H/R \sim (\sqrt{P_{\rm tot}/\rho})/V_{K}$ to turn Eq.~\ref{eqn:constraint} into a lower limit on $H/R$:
\begin{align}
\label{eqn:HRmin} \frac{H}{R} > \left( \frac{3 \dot{m}}{\alpha \Delta\, t_{S}} \right)^{1/2} \frac{R^{3/4}}{(G M_{\rm BH})^{1/4}} \sim 0.1 \frac{\dot{m}^{1/2} r_{0.1}^{3/4}}{\alpha_{0.1}^{1/2} \Delta_{0.01}^{1/2} m_{7}^{1/4}} \ .
\end{align}
In the last equality we define $\alpha_{0.1} \equiv \alpha/0.1$, $\Delta_{0.01} \equiv \Delta / 0.01$, typical values for many models for $\alpha$ and observational constraints on $\Delta$. 
Because $M_{\rm gas} \sim 2\pi\,R^{2}\,\Sigma$ and $\Sigma \sim 2\,\rho\,H$ in terms of the midplane gas density $\rho = \rho_{\rm mid}$ of the disk, we immediately obtain a corresponding upper limit to the (volume-averaged) midplane density: 
\begin{align}
\label{eqn:rhomax} \rho_{\rm mid} &< \frac{G^{1/4} M_{\rm BH}^{5/4} t_{S}^{1/2} \alpha^{1/2} \Delta^{3/2}} {4 \sqrt{3} \pi R^{15/4} \dot{m}^{1/2}} \\ 
\nonumber &\lesssim  \frac{5\times 10^{9}\,m_{p}}{\rm cm^{3}} \left( \frac{ m_{7}^{5/4} \alpha_{0.1}^{1/2} \Delta_{0.01}^{3/2}} { \dot{m}^{1/2} r_{0.1}^{15/4}} \right)\ .
\end{align}

Notably, in the classic ``$\alpha$-disk'' model where $\alpha$ is a constant and the pressure is primarily thermal, $\Pi \sim \alpha P_{\rm tot} \sim (0.01-0.1)\,P_{\rm thermal} \sim (0.01-0.1) (\rho/m_{p})\,k_{B} T_{\rm gas}$ is (by assumption) dominated by thermal gas pressure in the outer disk, and the ratio $\Pi/\rho \propto T_{\rm gas}$ depends \textit{only} on the gas temperature. But the gas temperature in such models is set by the accretion rate itself -- thus there is a robust constraint which does not allow any tunable parameters to resolve the key observational tensions we will discuss below.

\section{Observations}
\label{sec:obs}

Here we briefly review the compiled observations used to compare the models here, shown in Fig.~\ref{fig:vc.obs} and listed in Table~\ref{tbl:compilation}.
We attempt to compile observations from a wide variety of objects using a number of different techniques, for several reasons. First, these allow us to probe different size scales and regimes of BH mass/luminosity. Second, they provide a mutual consistency check that there are not large observational systematics biasing the conclusions. And third, they suggest that the results are representative of the general quasar population, not just populations which can be followed-up with a specific method. As we further discuss in \S~\ref{sec:conclusions}, the qualitative agreement of these different techniques, in samples spanning a range of BH mass and accretion rate, plus the fact that these samples do not appear to systematically deviate from ``typical'' AGN at similar luminosities in their continuum SED shapes (or other properties related directly to the accretion disk) suggest that they should be at least plausibly representative of the larger population. 

The objects surveyed span BH masses $\sim 10^{6}-10^{9}\,M_{\odot}$, and accretion rates/luminosities $L_{\rm bol} / L_{\rm Edd} \sim \dot{m} \sim 10^{-3} - 5$, with measurements at radii $R \sim 0.003 - 1$\,pc ($\sim 4-5000\,$ld, and $\sim 100 - 10^{7.5}\,R_{G}$ where $R_{G} \equiv G M_{\rm BH}/c^{2}$), though the majority (and the most constraining cases) tend to reside around $M_{\rm BH} \sim 10^{7}-10^{8}\,M_{\odot}$, $\dot{m} \sim 0.1-1$, $r \sim 0.01-0.3\,$pc ($10-1000$\,ld, $\sim 300-10^{5.5}\,R_{G}$). In the sections below, we describe the model predictions for e.g.\ thermal, magnetic, turbulent, and radiation-pressure supported disks for the quantities $V_{c}/V_{\rm K}$ and $\partial \ln V_{c} / \partial \ln R$ shown in Fig.~\ref{fig:vc.obs}: these can be scaled for each model to the exact value of $m_{7}$, $\dot{m}$, and $R$ for each observation, and we discuss whether there are any other parameters in those specific models (like $Q$ or $\alpha$) that could improve agreement. We use the individual values in Table~\ref{tbl:compilation} for all quantitative comparisons and computation of the limits to e.g.\ $H/R$, $\rho$, and the Toomre parameter $Q \equiv c_{s,\,{\rm eff}}\,\Omega / (\pi G \Sigma_{\rm gas})$ (\S~\ref{sec:theory}) and discuss them below, but for the sake of legibility in Fig.~\ref{fig:vc.obs}, we only plot reference model predictions for two representative variants of each of the four disk pressure assumptions ($\dot{m} \sim 0.1$ and $\sim 1$, spanning the range of the most interesting observations).\footnote{There are a few systems in Table~\ref{tbl:compilation}, namely NGC 4258, NGC 3783, J1339+1310, and J1206+4332, whose combination of BH mass, $\dot{m}$, and radii measured make them less-constraining for separating the models here at present.}

\begin{footnotesize}
\ctable[caption={{\normalsize Systems With Plotted Disk-Mass Constraints}.\label{tbl:compilation}},center,star]{lccccl}{
\tnote[ ]{For each AGN, we quote BH mass $m_{7} \equiv M_{\rm BH}/10^{7}\,M_{\odot}$, accretion rate $\dot{m}= 0.1\,\dot{M}\,c^{2}/L_{\rm Edd}$, range of radii of the measured kinematic constraints, method, and reference from which the constraints are taken.}}{
\hline
Name & $m_{7}$ & $\dot{m}$ & $R$ [${\rm pc}$] & Method & Reference \\
\hline
NGC 6323 & 1.00 & 1.4 & 0.14-0.31 & Maser & \citet{kuo:2018.maser.bh.masses.much.closer.to.keplerian.than.some.have.argued.percent.deviations.only} \\
UGC 3789 & 1.01 & 0.1 & 0.07-0.20 &  Maser & \citet{kuo:2018.maser.bh.masses.much.closer.to.keplerian.than.some.have.argued.percent.deviations.only} \\ % mdot ~ 0.03-0.15
NGC 6264 & 2.66 & 0.1 & 0.26-0.48 & Maser & \citet{kuo:2018.maser.bh.masses.much.closer.to.keplerian.than.some.have.argued.percent.deviations.only} \\ % mdot ~0.03-0.16
NGC 5765b & 4.87 & 0.04 & 0.30-1.3 & Maser & \citet{kuo:2018.maser.bh.masses.much.closer.to.keplerian.than.some.have.argued.percent.deviations.only} \\
NGC 2960 (Mrk 1419) & 1.18 & 0.008 & 0.12-0.31 & Maser & \citet{kuo:2018.maser.bh.masses.much.closer.to.keplerian.than.some.have.argued.percent.deviations.only} \\
NGC 4258 & 3.83 & 0.0004 & 0.11-0.30 & Maser & \citet{kuo:2018.maser.bh.masses.much.closer.to.keplerian.than.some.have.argued.percent.deviations.only} \\
NGC 2273 & 0.75 & 0.15 & 0.03-0.09 & Maser & \citet{kuo:2011.maser.bh.masses} \\ 
J0437+2456 & 0.29 & 0.01 & 0.04-0.13 & Maser & \citet{gao:2017.agn.masers.disk.mass.constraints.implicit} \\
ESO 558-G009 & 1.7 & 0.008 & 0.20-0.47 & Maser & \citet{gao:2017.agn.masers.disk.mass.constraints.implicit} \\ % not needed in final, too low-mdot
NGC 5495 & 1.1 & 0.004 & 0.10-0.30 & Maser & \citet{gao:2017.agn.masers.disk.mass.constraints.implicit} \\ % not needed in final, too low-mdot
%Mrk 1029 & 0.19 & --- & 0.23-0.45 & M & \citet{gao:2017.agn.masers.disk.mass.constraints.implicit} \\ too obscured to use
%NGC 1194 & --- & --- & --- & M & \citet{kuo:2011.maser.bh.masses} \\ % not used in final plot, not as constraining (just larger error bar)
%NGC 4388 & 0.85 & --- & --- & M & \citet{kuo:2011.maser.bh.masses} \\ % not used in final plot, not as constraining (just larger error bar - just a couple points)
\hline
Circinus & 0.17 & 0.18 & $<0.27$ & Imaging(+Maser) & \citet{izumi:2023.imaging.nuclear.gas.disk.circinus.accretion.rate} \\
\hline
Mrk 110 & 1.9 & 0.28 & 0.003-0.03 & Reverberation & \citet{Kollatschny2001} \\
Mrk 817 & 3.8 & 0.19 & 0.01-0.04 & Reverberation & \citet{Lu2021} \\
Mrk 509 & 10 & 0.16 & 0.03-0.07 & Reverberation & \citet{Zu2011} \\
3C 390.3 & 51 & 0.07 & 0.02-0.05 & Reverberation & \citet{Dietrich2012} \\
\hline
PDS 456 & 17 & 4.6 & 0.01-0.26 & Interferometry & \citet{gravity:2024.blr.infrared.size.luminosity.relation.agn} \\ 
3C 273 & 26 & 1.0 & 0.03-0.13 & Interferometry & \citet{gravity:2018.sturm.blr.rotating.thick.disk} \\ 
IC 4329a & 1.4 & 0.72 & 0.004-0.01 & Interferometry & \citet{gravity:2024.blr.infrared.size.luminosity.relation.agn} \\ 
Mrk1239 & 3.0 & 0.60 & 0.004-0.05 & Interferometry & \citet{gravity:2024.blr.infrared.size.luminosity.relation.agn} \\ 
Mrk 509 & 10 & 0.16 & 0.007-0.16 & Interferometry & \citet{gravity:2024.blr.infrared.size.luminosity.relation.agn} \\ 
IRAS 09149-6206 & 12 & 0.1 & 0.01-0.07 & Interferometry & \citet{gravity:2020.resolved.blr.size.disk.inside.dust.sub} \\ 
NGC 3783 & 4.8 & 0.05 & 0.007-0.03 & Interferometry & \citet{gravity:2021.resolved.blr.disk.hot.dust.coronal.regions} \\ 
\hline
J1004+4112 & 1.0 & 3.8 & 0.003-0.02 & Microlensing & \citet{fian:2024.microlensing.response.mapping.keplerian.rotation.curve.in.qsos} \\
J1001+5027 & 70 & 1.1 & 0.003-0.02 & Microlensing & \citet{fian:2024.microlensing.response.mapping.keplerian.rotation.curve.in.qsos} \\
HE 1104-1805 & 74 & 1 & 0.003-0.02 & Microlensing & \citet{fian:2024.microlensing.response.mapping.keplerian.rotation.curve.in.qsos} \\
J1206+4332 & 42 & 0.1 & 0.003-0.02 & Microlensing & \citet{fian:2024.microlensing.response.mapping.keplerian.rotation.curve.in.qsos} \\
J1339+1310 & 40 & 0.15 & 0.003-0.02 & Microlensing & \citet{fian:2024.microlensing.response.mapping.keplerian.rotation.curve.in.qsos} \\
\hline\hline}
\end{footnotesize}

\subsection{Masers}
\label{sec:obs:masers}

In many AGN, maser emission has enabled observational mapping of the kinematics and rotation curve of gas at radii $\sim 0.01 - $\,a few pc, most often around $\sim 0.1-1$\,pc surrounding black holes with masses $\sim 10^{6}-10^{8}\,M_{\odot}$ \citep{miyoshi:1995.agn.maser.discovery,greenhill:1995.maser.disk.detection.agn,braatz:agn.maser.search,herrnstein:2005.agn.acc.disk.constraints.from.maser,henkel:agn.masers}. This emission comes from molecular gas at densities $\sim 10^{7}-10^{11}\,{\rm cm^{-3}}$ and temperatures $\sim 100-1000\,$K, at radii $\sim 0.01-1\,$pc (references above and \citealt{modjaz:2005.agn.maser.modeling.Bfield.constraints.favor.fluxfrozen.disks,kondratko:agn.masers,kondratko:agn.masers.2,kondratko:ngc3393.acc.disk.maser}). Maser rotation curves have been extensively studied and modeled with sophisticated approaches that forward-model the observations directly from tilted-ring type assumptions allowing for different orbital anisotropy, eccentricity, warps, clumpiness, and other details. 
A common conclusion from these studies is that almost all maser rotation curves are at least consistent with a Keplerian potential (i.e.\ set some upper limit to $\Delta$), and the most well-behaved and highest-signal-to-noise examples set upper limits $\Delta \lesssim 0.01$ -- i.e.\ reach percent-level sensitivity to deviations from a Keplerian potential \citep{moran:1995.maser.review,lasker:2016.maser.host.agn.bh.host.scalings,gao:2017.agn.masers.disk.mass.constraints.implicit,kuo:2018.maser.bh.masses.much.closer.to.keplerian.than.some.have.argued.percent.deviations.only,linzer:2022.virial.single.epoch.bh.masses.unreliable}.

Here we compare the masers modeled in detail in \citet{kuo:2011.maser.bh.masses,kuo:2018.maser.bh.masses.much.closer.to.keplerian.than.some.have.argued.percent.deviations.only,gao:2017.agn.masers.disk.mass.constraints.implicit}. Specifically we take from their best-fit models the range of allowed deviations from Keplerian, including both their best-fit ``BH+disk'' models with the disk as a systematic deviation and the $\pm 2\,\sigma$ residuals with respect to the best-fit Keplerian model defined by the data therein. These are generally considered ``high quality'' masers, the most useful for our modeling here. Masers with more irregular kinematics can still place upper limits on $\Delta$ and therefore have constraining power (see \S~\ref{sec:noncircular}), but less so. We do not attempt to re-fit the datasets ourselves, but restrict to those with published detailed mass modeling.

As discussed below (\S~\ref{sec:thermal:mag}), some maser lines allow strong constraints on the disk magnetic fields, via upper limits to Zeeman splitting. We compile and compare these as well, where relevant.

\subsection{Torus-Scale Molecular Gas Imaging}
\label{sec:obs:imaging}

In addition to maser measurements, ALMA has recently made it possible (in the nearest AGN) to obtain sub-pc resolved imaging and spectroscopy of neutral gas (molecular and atomic metal-lines). These allow for simultaneous kinematic constraints but also, where the kinematics are ambiguous or not sufficiently well-resolved, a direct census of the gas mass. In Circinus, a direct census of the total (technically neutral, but that is expected and predicted to dominate at these radii) gas mass from imaging gives $M_{\rm gas} \sim 3500-6100\,{\rm M}_{\odot}$ ($\sim 0.002-0.003\,{\rm M}_{\rm BH}$) at $r<0.27\,$pc  \citep{izumi:2023.imaging.nuclear.gas.disk.circinus.accretion.rate}. Obviously $M_{\rm disk} \lesssim M_{\rm gas}$ (some of the gas could be outside the accretion disk), so this directly measures $\Delta$ and $M_{\rm gas}(<R)$ at this point.

\subsection{Broad-Line Regions}
\label{sec:obs:blr}

The broad line region (BLR) clouds are believed to be clouds orbiting in the vicinity of the supermassive black hole, at distances of $\sim 10^{-3}$--$10^{-1}$ pc (e.g., \citealt{Peterson2006:BLR.review,hickox:2018.agn.obscuration.review}). Therefore, their observed velocities and distances from the supermassive black hole can be used to probe the gravitational potential well, and in particular, to examine whether the kinematics are consistent with Keplerian motions caused by a single central point source mass. There are three types of BLR observations that probe the kinematics and spatial extents of the BLR clouds which we attempt to compile here: Reverberation Mapping (RM; see review by \citealt{cackett:2021.reverberation.mapping.multiwavelength.review}); imaging that directly resolves the BLR using interferometry (GRAVITY; e.g., \citealt{gravity:2018.sturm.blr.rotating.thick.disk}); and kinematic microlensing responses (microlensing; \citealt{vernardos:2024.qso.microlensing.review}) which are described below. 

\subsubsection{Reverberation Mapping}
\label{sec:obs:blr:rm}
         
The BLR line fluxes are observed to vary in response to variations in the continuum emission with a short time delay (days to weeks; e.g., \citealt{Kaspi2000, Peterson2004}, and review by \citealt{cackett:2021.reverberation.mapping.multiwavelength.review}). This time delay is attributed to the light travel time across the BLR, and thus, represents the light-weighted distance of the line-emitting clouds from the continuum source. Since the continuum source, the accretion disk, is on much smaller scales of $\sim 10^{-5}$--$10^{-3}$ pc, the time delay is considered to represent the distance from the supermassive black hole. RM observations of several emission lines that trace gas at different ionization levels show that higher ionization lines, such as C IV and He II, have shorter time delays and broader Doppler widths, consistent with the idea that they trace gas that is orbiting closer to the supermassive black hole. In several well-studied systems such as NGC 5548, the different emission lines span a large-enough range in distance and show Doppler widths consistent with a $v \propto r^{-1/2}$ relation, the expected relation for Keplerian motions due to the supermassive black hole (e.g., \citealt{Peterson1999, Bentz2009}). 

Most RM campaigns focus on mapping a single emission line (e.g., H$\alpha$ or H$\beta$) to obtain the size of the BLR. To test whether the motions are consistent with Keplerian motion, RM observations of several different ionization emission lines are required to probe different distances from the supermassive black hole simultaneously. Since the continuum emission may vary significantly over timescales of months-years, and due to the BLR size-luminosity relation, the different emission lines have to be observed over the same period of time. In addition, Fig.~\ref{fig:vc} shows that the \citet{shakurasunyaev73} model is expected to diverge from Keplerian on scales of $\sim$0.03 pc. This implies that most BLR RM observations published in the literature, which target lower luminosity black holes with BLR sizes generally below 0.03 pc (e.g., \citealt{Kaspi2000, Peterson2004, Bentz2009}), cannot be used to test the models. BLR RM mapping of higher luminosity AGN require significantly longer observational campaigns, with very few published results for quasars (e.g., \citealt{Lira2018}). Nevertheless, a few higher luminosity cases have been observed with multiple emission lines (Mrk 817, 3C390.3, Mrk 509 and Mrk 110; see \citealt{Kollatschny2001, Zu2011, Dietrich2012, Lu2021}), and we show their derived velocities and distances in Fig.~\ref{fig:vc.obs}. In all but one case (Mrk 110), we use the line dispersion in the variable part of the spectrum to define the velocity, and the cross-correlation function centroid to define the time delay and thus the distance (see e.g., \citealt{Peterson2004}). Caution must still be taken when comparing these observations with model predictions, noting (i) the large uncertainties on the derived line kinematics, which also change with the ionizing luminosity during the campaign, and (ii) the assumption that each emission line represents a specific distance from the supermassive black hole. Finally, more recent, 2D RM campaigns suggest that the BLR motions in some sources are consistent with an outflow (see review by \citealt{cackett:2021.reverberation.mapping.multiwavelength.review}). Such analysis has only been applied to low-luminosity sources with BLR sizes lower than 0.03 pc, and thus they do not directly apply to Fig.~\ref{fig:vc.obs}.

\subsubsection{Near-Infrared Interferometry}
\label{sec:obs:blr:interferometry}

The second, more recent, method that probes the BLR kinematics and geometry is direct imaging of the BLR clouds using long baseline near-infrared interferometry (GRAVITY; \citealt{gravity:2018.sturm.blr.rotating.thick.disk}). GRAVITY, a recently deployed instrument at the Very Large Telescope Interferometer in Chile, is capable of reaching a spectro-astrometric precision of micro arc-seconds, allowing it to spatially resolve the BLR clouds. In their first paper, \citet{gravity:2018.sturm.blr.rotating.thick.disk} identified a spatial offset between the red and blue photo-centres of the broad Pa$\alpha$ line of the quasar 3C 273. The detected spatial offsets and velocity gradients imply that the BLR gas is orbiting around the supermassive black hole, with their data well-fitted by a BLR model of a thick disk of gravitationally bound gas orbiting the black hole. We present the spatial offsets and velocity gradients in Fig.~\ref{fig:vc.obs}. So far, GRAVITY had mapped the BLR of 7 objects in total, with derived BLR sizes between $\sim$10 light-days (0.0084 pc) to $\sim$300 light-days (0.25 pc; see \citealt{gravity:2020.resolved.blr.size.disk.inside.dust.sub, gravity:2021.resolved.blr.disk.hot.dust.coronal.regions, gravity:2024.blr.infrared.size.luminosity.relation.agn}). These observations are fitted with an elaborate model of the BLR that includes a distribution of non-interacting clouds in a Keplerian/inflow/outflow motion, with varying spatial distributions in the $r$ and $\theta$ directions, allowing the derivation of the black hole mass, the BLR mean size and thickness, and the BLR motion. The observations of five out of the seven sources are consistent with a rotating thick disk BLR, while the two others are consistent with outflow-dominated BLRs (Mrk 509 and PDS 456; \citealt{gravity:2024.blr.infrared.size.luminosity.relation.agn}). 

\subsubsection{Microlensing}
\label{sec:obs:blr:microlensing}

Third and most recently, \citet{fian:2024.microlensing.response.mapping.keplerian.rotation.curve.in.qsos} attempted to reconstruct the circular velocity profile of five luminous quasars (SDSS J1001+5027, J1004+4112, J1206+4332, J1339+1310, \&\ HE 1104-1805) from the microlensing response of different velocity components of the C IV and Si IV broad emission lines. Microlensing has been used to estimate the size of the AGN continuum-emitting and broad-line regions at different wavelengths, as the response function is smoothed and suppressed depending on the size of the emission region (see \citealt{vernardos:2024.qso.microlensing.review} for a review). \citet{fian:2024.microlensing.response.mapping.keplerian.rotation.curve.in.qsos} find that reconstructing the sizes of different velocity components leads to a remarkably close-to-Keplerian circular velocity curve, in either C IV or Si IV and in either the blue or red wing of either line, for all five systems they study at $\sim 5-20\,$light-days (below which they obtain primarily size upper limits). Note \citet{husemekers:2024.microlensing.kinematics.blr} argue that there is still non-trivial degeneracy in the models when an arbitrary distribution of broad-line emitter clouds plus microlensing objects is fitted, in particular that some (though not all) of the systems in \citet{fian:2024.microlensing.response.mapping.keplerian.rotation.curve.in.qsos} can be comparably or better fit by ``equatorial wind'' as compared to rotating disk models. However, we stress that our constraint is on the {\em gravitational potential}, not on the BLR velocity field. There is a long history of MHD-wind and outflow models for the BLR, going back to \citet{emmering:1992.blr.in.mag.winds,murray:1995.acc.disk.rad.winds} and explicitly tested against NGC 5548 observations in \citet{bottorff:1997.ngc5548.blr.mhd.wind.model.outflow}; more recent microlensing-era analyses  \citep{braibant:2017.modeling.blr.geometry.kinematics.from.microlensing,hutsemekers:2024.civ.bel.qso.region.size,husemekers:2024.microlensing.kinematics.blr} similarly fit equatorial-wind kinematics to BLR observations. Crucially, all of these models -- including \citet{emmering:1992.blr.in.mag.winds} and \citet{bottorff:1997.ngc5548.blr.mhd.wind.model.outflow} -- explicitly assume a Keplerian {\em potential} set by $M_{\rm BH}$ alone, and model non-Keplerian gas {\em kinematics} on top of it. Our constraint concerns the former, not the latter, and is therefore consistent with -- rather than in tension with -- the BLR outflow/wind literature. Concretely: (1) per \S~\ref{sec:noncircular}, the functional forms used in \citet{braibant:2017.modeling.blr.geometry.kinematics.from.microlensing} and \citet{hutsemekers:2024.civ.bel.qso.region.size,husemekers:2024.microlensing.kinematics.blr} inherit a strictly Keplerian potential from \citet{murray:1995.acc.disk.rad.winds}, with the gas in a disky-outflow {\em geometry}. And the deviations in the velocities which would appear if one introduced an additional disk mass to the potential have the \textit{opposite} sense of the observational residuals they argue the wind model fits (i.e.\ they make the fit significantly worse). And (2) given our simplistic comparisons, our conclusions are identical if we restrict to the set of ``cleaner'' systems best-fit by the rotating disk models in \citet{hutsemekers:2024.civ.bel.qso.region.size,husemekers:2024.microlensing.kinematics.blr}.

\begin{figure*}
	\centering
	\includegraphics[width=0.99\textwidth]{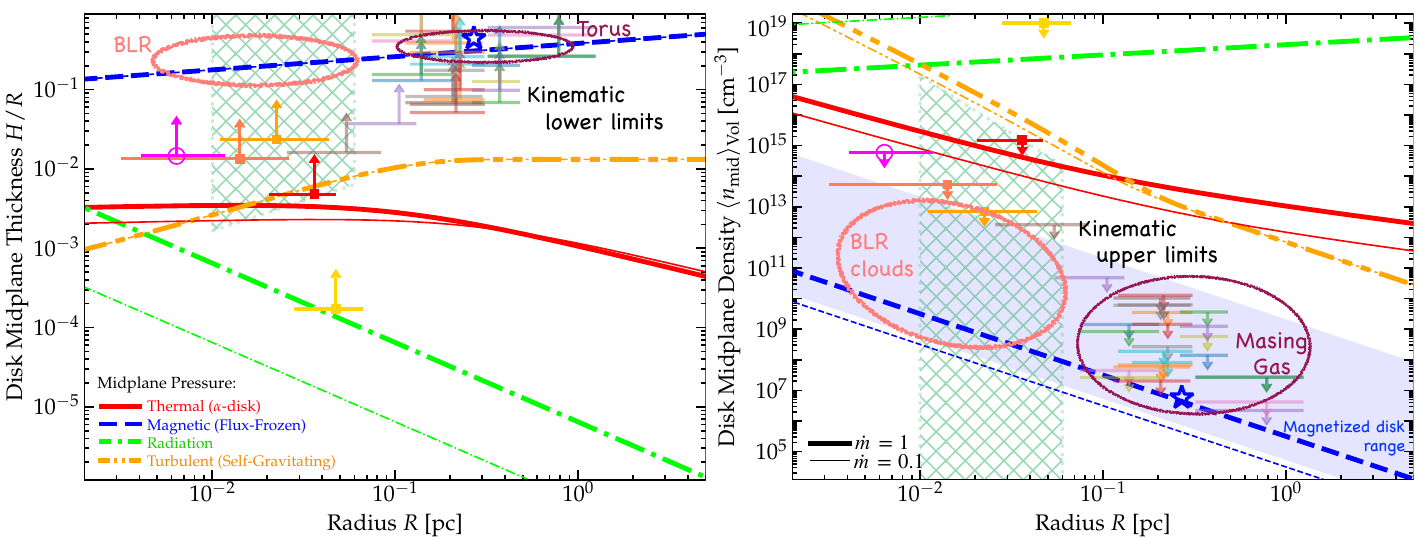} 
	\caption{Corresponding predictions (as Fig.~\ref{fig:vc}) and lower/upper limits (Eqs.~\ref{eqn:HRmin}-\ref{eqn:rhomax}) \textit{purely from the kinematic constraints} in Fig.~\ref{fig:vc.obs} on the central accretion disk scale-height $H/R$ (\textit{top}) and volume-averaged midplane density $n_{\rm mid}$ (\textit{bottom}). These constraints do not depend on the tracers being ``part of'' the disk, only that they feel the gravitational potential of the disk (and therefore constrain its mass). 
	For the magnetically-dominated disk, since the disks are predicted to be highly multi-phase, we show the expected range of densities (\textit{shaded}).
	We also show (\textit{ellipses}) the approximate typical height and radius range of the BLR and ``torus'' regions, from typical covering factors and resolved imaging, as well as the typical densities of BLR clouds and of masing gas (from spectral modeling). 
	The kinematic constraints clearly favor the magnetized disk model at $\gtrsim 0.003\,$pc. The limits on $H/R$ and density imply that the BLR, masers, and torus do not ``sit above'' a much denser, thinner disk containing most of the mass, but rather are fundamentally part of the same structure.
	\label{fig:HR.rho}}
\end{figure*}

\subsection{Constraints on the Gravitational Potential versus non-Circular Motions}
\label{sec:noncircular}

It is important to distinguish between the \textit{circular velocity} $V_{c}$ (Eq.~\ref{eqn:vc}), which by definition is a statement about the potential and enclosed mass, and the \textit{rotation velocity} $V_{\rm rot}$ or line-of-sight projected mean velocity $V_{\rm los}$ of gas at some impact parameter $b$ in projection from the SMBH. Gas can exhibit non-circular or non-Keplerian motions ($V_{\rm rot} \ne V_{c}$) in a strictly Keplerian potential, owing to e.g.\ inflows/outflows, eccentric/radial motion, or clumpiness/turbulence. Detailed modeling can often disentangle these \citep[see discussion in][]{kuo:2018.maser.bh.masses.much.closer.to.keplerian.than.some.have.argued.percent.deviations.only,gallimore:2023.ngc.1068.no.massive.acc.disk.just.eccentric.warp}, and various studies have found that almost all maser systems with apparently non-Keplerian kinematics appear to be non-rotating \citep[often impacted by outflows and/or jets; see references above and][]{lonsdale:vlbi.agn.cores,henkel:agn.masers,kuo:2020.agn.masers.impacted.by.outflows,panessa:2020.maser.agn.hardxr.selection.categories}. In these cases, what is measured still provides an upper limit to $\Delta$, even if the magnitude of the fractional contribution of e.g.\ non-circular motion versus variation in the shape of the potential is unable to be determined. That, in turn, means that we still obtain a lower limit to $P_{\rm tot}/\rho$ -- i.e.\ even these measurements are still able to give us a useful constraint. Of course, larger/less-strongly constrained upper limits to $\Delta$ are less constraining for our purposes. Similarly, at large radii (e.g.\ $\gtrsim$\,pc), there could be additional contributions to the potential or enclosed total mass $M_{\rm enc}$ from material besides the accretion disk or BH, including other (non-accreting) gas, stars, stellar-mass black holes, or dark matter. But again, given some observed rotation curve and upper limit on $\Delta$, if there is any such contribution, then the accretion disk mass must be even lower than the limit derived above from $\Delta$, and therefore $P_{\rm tot}/\rho$ would have to be even larger. So again because our constraints rely only on having a lower limit to $P_{\rm tot}/\rho$, they are robust to these effects (if there is significant contribution to the potential from stars or dark matter or black holes, it would only strengthen our conclusions). But in either case the most interesting/useful constraints will come from observational cases which are closest-to-Keplerian, where the degeneracies of modeling/inferring $V_{c}$ are minimized and where $\Delta \lesssim 0.01$ \citep{kuo:2018.maser.bh.masses.much.closer.to.keplerian.than.some.have.argued.percent.deviations.only}.

To our knowledge, there are only two maser cases with acceptable fits\footnote{We exclude cases like NGC 7738, recently argued in \citet{ito:2025.maser.kinematics.7738.but.did.fits.to.vr.not.rv.high.chi2} to have $M_{\rm disk} \gg M_{\rm BH}$, as the best-fit model is statistically a poor fit to the data (reduced $\chi^{2} / \nu \equiv [N_{i}-N_{\rm model}]^{-1}\,\sum_{i} (x_{i} - x_{i}^{\rm model}[v^{\rm los}_{i}])^{2} / \sigma[x]_{i}^{2} \sim 3.9$, or as-fit $\chi^{2}_{v} / \nu \equiv [N_{i}-N_{\rm model}]^{-1}\,\sum_{i} (v^{\rm los}_{i} - v_{i}^{\rm model}[x_{i}])^{2} / \sigma[v^{\rm los}]_{i}^{2} \sim 160$), indicating the fit is driven by more complex kinematics. Moreover that study utilized a linear $\chi^{2}$ fit to $V_{\rm los}(x_{i})$, despite the dominant errors being in the maser positions $x_{i}$; re-fitting their published data with a simple 2D maximum-likelihood model instead of a linear $\chi^{2}$ gives a best-fit rotation curve slope completely consistent with Keplerian ($\partial \ln{V_{\rm rot}}/\partial \ln{R} \rightarrow -0.57 \pm 0.11$ with $\chi^{2}/\nu\rightarrow 0.995$, as compared to the linear $\chi^{2}$ fit $\partial \ln{V_{\rm rot}}/\partial \ln{R} \rightarrow -0.14 \pm 0.04$ with $\chi^{2}/\nu\rightarrow 3.9-160$).} which have been specifically argued by some as showing positive evidence of a non-Keplerian \textit{potential}, NGC 3079 and 1068. We will discuss these specifically below in our comparisons and show that even taking the claimed detections at face value, the implied disk masses still rule out most disk models. But we would more generally argue these should still be treated as upper limits. For NGC 3079, the claim is fairly tentative -- \citet{kondratko:2005.3079.selfgrav.disk.mass.masers} argue for some disk mass at $<0.7$\,pc, with a limit $M_{\rm disk}^{3079} \lesssim 1\times10^{6}\,M_{\odot}$ (with much larger uncertainties beyond that radius), but this system is well-known to have a bipolar jet/wind impacting the maser emission region directly, potentially producing irregular kinematics (the wind correction to the gravitational motions is order-unity). For NGC 1068 ($M_{\rm BH} \sim 0.8-1.7 \times 10^{7} M_{\odot}$, $\dot{m} \sim 0.2$), the most extreme claim of non-Keplerian motion comes from \citet{lodato:2003.ngc.1068.agn.massive.acc.disk.some.evidence.rules.out.ss73}, who argue for an ``additional'' (non-BH) mass $M^{1068}_{\rm disk} \lesssim M^{1068}_{\rm non-BH} \sim 8\times10^{6} {\rm M}_{\odot} \sim M_{\rm BH}$. But these authors still find an upper limit for the \textit{slope} of the rotation curve $V_{c} \propto r^{\zeta_{V}}$ with $\zeta_{V} \le 0$ (i.e.\ flat-or-declining $V_{c}$), which we will show rules out most thermal/radiation-pressure dominated disk models. And as they themselves note, there is clear evidence for at least some of this mass coming from stars in the nuclear star cluster (extrapolation of whose mass profile could account for all of the apparent rotation curve derivation), hence the upper limit  $M^{1068}_{\rm disk} \lesssim M^{1068}_{\rm non-BH}$. Moreover, reanalysis with higher-resolution and more extensive datasets by \citet{gallimore:2023.ngc.1068.no.massive.acc.disk.just.eccentric.warp} has argued that there is no evidence for non-Keplerian circular velocities (the feature in the observed velocities being driven by eccentric motions), and those authors set an updated upper limit $M^{1068}_{\rm disk} < 10^{5}\,{\rm M}_{\odot}$.

\section{Constraints on Different Pressures}
\label{sec:pressures}

In Figs.~\ref{fig:vc.obs}-\ref{fig:HR.rho}, we compare the observed direct dynamical constraints on $V_{c}/V_{\rm K}$ (i.e.\ $M_{\rm disk}/M_{\rm BH}$), and correspondingly $H/R$ and $\rho$, to different disk models. We immediately see that dynamical measurements rule out standard models of thermal-pressure, radiation-pressure, cosmic-ray pressure, and turbulence-only (with $Q\lesssim 1$) dominated disks, leaving magnetic-pressure dominated disks as the only viable option, at radii $\sim 0.003 - 3\,$pc ($\sim 4-5000$\,light-days, or $\sim 300 - 10^{7}\,R_{\rm G}$). In the following sections, we discuss this in more detail for each model in turn, and ask whether there is any adjustable model parameter which could alleviate these constraints. We further go into details on additional constraints and checks for each model (e.g.\ upper limits on the observed disk temperatures, luminosities, energetics, and magnetic field strengths), to support these conclusions.

Before turning to each pressure source in turn, we emphasize a logical point which will be central to the discussion below. As derived in \S~\ref{sec:theory}, any self-consistent accretion disk model must satisfy a {\em lower} limit on the specific stress-to-mass ratio $\Pi_{\rm eff}/\rho$, which in turn cannot exceed $P_{\rm tot}/\rho$ of whatever pressure source supports the disk (Eq.~\ref{eqn:constraint}). The outer-disk self-gravity problem has a long history: the fact that standard $\alpha$-disks become self-gravitating and massive at $R \gtrsim 10^{-2}\,$pc was already noted by \citet{shlosman:inefficient.viscosities} and \citet{goodman:qso.disk.selfgrav}, and a number of solutions have been proposed -- including marginal self-regulation at $Q\sim1$ \citep{paczynski:1978.selfgrav.disk,gammie:2001.cooling.in.keplerian.disks,sirko:qso.seds.from.selfgrav.disks,thompson:rad.pressure}, fragmentation and wind/radiation-pressure ablation \citep{shlosman:inefficient.viscosities,goodman:qso.disk.selfgrav}, bar-driven and gravitational-torque-driven fueling \citep{shlosman:bars.within.bars,begelman:direct.bh.collapse.w.turbulence,choi:2013.direct.collapse.smbh.bar.drainage.self.grav.disk.sims,hopkins:zoom.sims,hopkins:inflow.analytics,hopkins:qso.stellar.fb.together,hopkins:m31.disk,angles.alcazar:grav.torque.accretion.cosmo.sim.implications,daa:20.hyperrefinement.bh.growth}, and more recently the hyper-magnetized/flux-frozen models we focus on here. Many of these mechanisms can in principle reduce the outer disk mass at a given radius, thereby helping to reconcile the predicted rotation curve with observations. What is new here is the synthesis of different models, the collection and comparison with new observational constraints to turn this from a theoretical challenge/question into a direct observational constraint on disk properties, and the important (and often overlooked) point that the kinematic constraints in Figs.~\ref{fig:vc.obs}-\ref{fig:HR.rho} translate directly to a dynamical constraint on the {\em pressure source} of whatever disk remains. Reducing the disk mass in steady-state necessarily requires fast depletion times, and those in turn immediately imply that the remnant disk cannot be thermal-, radiation-, or cosmic-ray-pressure-dominated (at the radii $\gtrsim 10^{-3}$\,pc of interest), because those pressures cannot supply the required $\Pi_{\rm eff}/\rho$ without grossly violating independent observational constraints on disk luminosity and temperature. The remainder of this section demonstrates this systematically.

\subsection{Thermal-Pressure Dominated Disks (SS73)}
\label{sec:thermal}

\subsubsection{The Standard $\alpha$-Disk Is Immediately Ruled Out By Dynamical Constraints}
\label{sec:thermal:default}

If thermal pressure dominates, $P_{\rm tot} \approx P_{\rm thermal} = n k_{B} T = \rho c_{s}^{2}$. For disks with intermediate accretion rates ($0.001 \lesssim \dot{m} \lesssim 1$), the self-consistent accretion disk solution in this limit is that of SS73.\footnote{Note that we will ignore radiatively inefficient, very-low $\dot{m}$ optically-thin ADAF type solutions \citep[e.g.][]{yuan:2014.hot.accretion.flows.review} as we are interested in AGN accreting at modest luminosities. The classic super-Eddington ``slim-disk'' type extensions of SS73 \citep{paczynsky.wiita:1980.slim.disk,abramowicz:1988.slim.disks} fall into the radiation-pressure dominated category, which we discuss below.} It is straightforward to integrate that model, noting that by definition at the radii of interest we are well outside their regime ``(a)'' (radiation-pressure dominated), so we consider their regimes ``(b)'' (partially ionized, so scattering-opacity dominated) or ``(c)'' (more fully-neutral) therein, to obtain the predicted disk mass correction: 
\begin{align}
& \frac{M^{\rm therm}_{\rm gas}(<R)}{M_{\rm BH}} \sim \psi_{\rm therm} \left(1 + \psi_{\rm therm}^{1/4} \right)  \\ 
\nonumber & \ \ \ \ \ \psi_{\rm therm} \sim 1.2\,\frac{\dot{m}^{7/10} r_{0.1}^{7/5}}{m_{7}^{1/5} \alpha_{0.1}^{4/5}} {\rm MAX} {\Bigl[} 1 \ , \ 
0.8\left(\frac{m_{7}}{r_{0.1}}\right)^{0.15} 
{\Bigr]} 
\end{align}
where the MAX in $\psi$ reflects the regime (b)-(c) transition. 

We also immediately obtain the related predictions: 
\begin{align}
\frac{H^{\rm therm}}{R} &\sim 0.004 \, 
{\rm MIN} \left[ 
\frac{\dot{m}^{1/5} r_{0.1}^{1/20}}{m_{7}^{3/20} \alpha_{0.1}^{1/10}} \ ,\ \frac{\alpha_{0.1}^{1/4}}{(m_{7} \dot{m})^{1/16} r_{0.1}^{9/16}}
\right] \\
V_{c,\,{\rm disk}}^{\rm therm} &\sim 740\,{\rm km\,s^{-1}}\,m_{7}^{3/8} \dot{m}^{7/16} \alpha_{0.1}^{-1/2} r_{0.1}^{3/8} \\
T^{\rm therm} &\sim 2100{\rm K} \times {\rm MAX} \left[ 
\frac{m_{7}^{11/20} \dot{m}^{3/10}}{r_{100}^{3/4} \alpha_{0.1}^{1/5}}
\ , \ 
\frac{m_{7}^{5/8} \dot{m}^{5/8}}{r_{100}^{3/8} \alpha_{0.1}^{1/2}}
 \right]\\ 
n^{\rm therm} &\sim 1.5 \times 10^{13}\,{\rm cm^{-3}} \times \\ 
\nonumber & {\rm MAX}\left[ \frac{m_{7}^{47/40}\,\dot{m}^{11/20}}{r_{100}^{15/8} \alpha_{0.1}^{7/10}} \ , \ 
\frac{m_{7}^{13/16} \dot{m}^{13/16}}{r_{0.1}^{11/16}\alpha_{0.1}^{5/4}} \right]
\end{align}
where $V_{c,\,{\rm disk}}^{\rm therm}$ refers to the contribution to $V_{c}$ just from the disk self-gravity, and $H^{\rm therm}$, $T^{\rm therm}$, $n^{\rm therm}$ are the predicted scale-height/density/temperature from this model.
We scale to the canonical $\alpha \sim 0.1$ for an SS73 disk though even $\alpha \sim 1$ (while already ruled out from other observational and theoretical constraints; see \citealt{abramowicz:accretion.theory.review}) would not change our conclusions.

These scalings are immediately ruled out by the constraints in Figs.~\ref{fig:vc.obs}-\ref{fig:HR.rho} (rescaling for different $\alpha<1$, $m_{7}$, $\dot{m}$ makes no difference, generally only making the discrepancies larger, when we account for individual variations in these quantities between observed systems). 
In short, the model clearly predicts circular velocities exceeding Keplerian motions due to the black hole, with a rising rotation curve, i.e.\ $\Delta \gtrsim 1$ ($\gg 1$, at radii approaching $\sim 1\,$pc), $H/R$ much smaller than allowed by Eq.~\ref{eqn:HRmin}, as well as temperatures a factor $\sim 20$ larger than would permit molecular emission and gas densities $\sim 4-6$ orders of magnitude larger than the masing gas. And it predicts the disk is optically-thick to its own maser emission (with $\tau \gg 1$ even at $\gtrsim$\,pc radii), so the emission observed cannot possibly come from the disk. Even if somehow that gas were ``not part of the disk'' (in e.g.\ some ``skin'' above the disk, which somehow would have to be much colder while still being less dense, exactly the opposite of the SS73 predictions), the dynamics of the disk immediately strongly rule out such a model. 

Not only does the thermal-pressure dominated model predict much too large a disk mass compared to what is allowed, but it also predicts a \textit{rising} rotation curve, with $V_{c} \propto r^{+3/8}$ \textit{increasing} with radius, much more steeply than even the most extreme allowed ``disk-like'' components observed \citep{lodato:2003.ngc.1068.agn.massive.acc.disk.some.evidence.rules.out.ss73,kondratko:2005.3079.selfgrav.disk.mass.masers,kuo:2018.maser.bh.masses.much.closer.to.keplerian.than.some.have.argued.percent.deviations.only}. 

To be specific, consider the claims of a ``detection'' (albeit indirect) of non-zero disk mass within the BHROI. In Circinus, a direct census of the total (neutral, but that is expected and predicted to dominate at these radii) gas mass from imagining gives $M_{\rm disk}^{\rm Circ} \sim 3500-6100\,{\rm M}_{\odot}$ ($\sim 0.002-0.003\,{\rm M}_{\rm BH}$) at $r<0.27\,$pc  \citep{izumi:2023.imaging.nuclear.gas.disk.circinus.accretion.rate}. For its BH mass and accretion rate (tabulated below, but well into the regime $\dot{m} \sim 0.1$ where these predictions and model comparisons should apply), the prediction from this model is $M_{\rm disk}^{\rm therm} \sim 5.2 \times 10^{6} {\rm M}_{\odot}$ -- one thousand times larger than observed. Next consider NGC 1068  ($M_{\rm BH} \sim 0.8-1.7 \times 10^{7} M_{\odot}$, $\dot{m} \sim 0.2$), perhaps the most well-studied BH where some have claimed maser evidence for a non-Keplerian rotation curve from disk self-gravity. The most extreme such claim from \citet{lodato:2003.ngc.1068.agn.massive.acc.disk.some.evidence.rules.out.ss73} argues that the upper limit for the \textit{slope} of the rotation curve is flat, i.e.\ $V_{c} \propto r^{\zeta_{V}}$ with $\zeta_{V} \le 0$, while the model here predicts $+3/8$, clearly ruled out. And their inferred disk mass is $M^{1068}_{\rm disk} \lesssim 8\times10^{6} {\rm M}_{\odot}$: but for these parameters the thermal-pressure dominated models predict $M_{\rm disk}^{\rm therm} \approx 4\times 10^{8}\,{\rm M}_{\odot}$, a factor 100 larger than observationally claimed. And as noted in \citet{lodato:2003.ngc.1068.agn.massive.acc.disk.some.evidence.rules.out.ss73}, there is clear evidence for at least some of this ``disk'' mass coming from stars in the nuclear star cluster (extrapolation of whose mass profile could account for all of the apparent rotation curve derivation), hence the upper limit here. Moreover for this particular case, reanalysis with higher-resolution and more extensive datasets by \citet{gallimore:2023.ngc.1068.no.massive.acc.disk.just.eccentric.warp} has argued that there is no evidence for non-Keplerian circular velocities (the feature in the observed velocities being driven by eccentric motions), and those authors set an updated upper limit $M^{1068}_{\rm disk} < 10^{5}\,{\rm M}_{\odot}$, which is more than 1000 times smaller than the thermal disk predictions. NGC 3079 may also have a massive disk \citep{kondratko:2005.3079.selfgrav.disk.mass.masers}, but also is well-known to have a bipolar jet/wind impacting the maser emission region directly, potentially producing irregular kinematics (the wind correction to the gravitational motions is order-unity). Still \citet{kondratko:2005.3079.selfgrav.disk.mass.masers} argue for some disk mass at $<0.7$\,pc, with a limit $M_{\rm disk}^{3079} \lesssim 1\times10^{6}\,M_{\odot}$ (with much larger uncertainties beyond that radius). But again, the thermal disk model for the parameters of 3079 predicts $M_{\rm disk}^{\rm therm} \sim 1.3 \times 10^{8} {\rm M}_{\odot}$ at this radius, a factor $>100$ times larger than allowed. And in essentially all other cases, the upper limits to the disk mass are much lower (more like our default-scaled $\Delta \sim M_{\rm gas}/M_{\rm BH} \lesssim 0.01$; \citealt{kuo:2018.maser.bh.masses.much.closer.to.keplerian.than.some.have.argued.percent.deviations.only}). Thus even the most extreme claims of detected self-gravitating accretion disks, taken at face value, are multiple orders-of-magnitude less massive than the disk masses predicted if the disks were thermal-pressure dominated or SS73-like.

We note that the asymmetric-drift / pressure-support correction to $V_{c}$ in this model is $\delta V_{c} / V_{c} \sim (c_{s,\,{\rm eff}}/V_{K})^{2} \sim (H/R)^{2}$, with $H/R \lesssim 10^{-2}$ predicted (see above), so this correction is $\lesssim 10^{-4}$ -- completely negligible on the scale of Fig.~\ref{fig:vc.obs} and smaller than the uncertainties of the models plotted \citep[as emphasized in the context of gas-dynamical SMBH mass measurements in][]{walsh:2013.m87.gas.dynamics.mbh.constraint}. This is qualitatively different from the magnetic case (\S~\ref{sec:magnetic:pressurefx}), where $H/R \sim 0.1$ makes the correction potentially percent-level and we treat it explicitly.

\subsubsection{Any Other Thermal-Pressure-Dominated Disk Is Also Ruled Out}
\label{sec:thermal:arbitrary}

What if we completely ignored the physics used to derive the classic $\alpha$-disk model above (i.e.\ physically-self-consistent opacities and temperature structure, the assumption that the disk is optically thick, even basic physical constraints like energy conservation), and simply allowed the disk to have $P_{\rm tot} \approx P_{\rm thermal}$ with any temperature (i.e.\ simply ``fit'' a temperature to be consistent with Eq.~\ref{eqn:constraint})? 
In this limit, our constraint on $\Delta$ through Eq.~\ref{eqn:constraint} immediately translates to a lower bound on $T$: 
\begin{align}
\label{eqn:thermal.Tmin} T^{\rm therm}(R) 
\gtrsim 10^{6} {\rm K} \frac{\dot{m} m_{7}^{1/2} r_{0.1}^{1/2} \mu_{m}}{\alpha_{0.1} \Delta_{0.01}} 
\end{align}
This can be immediately ruled out for several reasons. Theoretically, (1) it is impossible to sustain -- at these temperatures the disk would cool far too efficiently and the cooling rate at these radii would be orders-of-magnitude larger than the gravitational energy flux (this is just the statement that the disk could \textit{not} be self-consistent); and (2) at these temperatures and radii one could not avoid radiation pressure becoming larger than thermal. But more importantly, purely observationally, we know (1) this temperature is far too hot to allow maser emission to exist; (2) this is also far hotter than the thermal emission temperature from the BH (from the spectrum of the big blue bump); and (3) this would necessarily imply an unphysically large luminosity from the gas at these radii. For the latter, taking $L(R) \approx 4\pi\sigma_{B}\,T_{\rm eff}^{4}$ as the luminosity from each annulus $R$, consistent with the regime that the disk would have to be in here, and comparing this to the Eddington limit, Eq.~\ref{eqn:thermal.Tmin} implies: 
\begin{align}
\label{eqn:thermal.Lmin} \frac{L^{\rm therm}(R)}{L_{\rm Edd}} \gtrsim 10^{11} m_{7} \left( \frac{\dot{m} r_{0.1} \mu_{m} }{\alpha_{0.1} \Delta_{0.01}}\right)^{4}
\end{align}
i.e.\ more than 10 orders of magnitude larger than observed! Moreover $L(R)$ increases as $R^{4}$, implying almost all the observed quasar emission comes from $>$\,pc scales, which is clearly ruled out by both variability and direct imaging constraints. 

Thus completely independent of assumptions for $\alpha$ or details of the accretion disk structure, we can immediately observationally rule out that the disks \textit{at the radii of maser/outer-BLR/dust emission} are thermal-pressure dominated.

\begin{footnotesize}
\ctable[caption={{\normalsize Direct Maser Constraints on in-plane $B$}.\label{tbl:Bfields}},center,star]{lccccl}{
\tnote[ ]{For each AGN, we quote BH mass $m_{7}$, accretion rate $\dot{m}$, claimed upper limit  $B^{\rm lim}_{\|}$ ($3\sigma$ or limit available from reference) to  in-plane $B_{\phi}$ or $| B_{\|} |$ in mG, and radius $R$ in pc where the upper limit is measured, from the given references.}}{
\hline
Name & $m_{7}$ & $\dot{m}$ & $B^{\rm lim}_{\|}$ [mG] & $R^{\rm lim}$ [pc] & Reference \\
\hline
NGC 3079 & 0.2 & 0.1-0.5 & 33 & 0.64 & \citet{vlemmings:2007.maser.b.limits} \\
NGC 4258 & 4 & $10^{-(3.5-4)}$ & 30-90 &  0.14-0.27 &  \citet{modjaz:2005.agn.maser.modeling.Bfield.constraints.favor.fluxfrozen.disks} \\
NGC 1194 & 6.5 & 0.02 &  100 & 0.6 & \citet{pesce:2015.agn.maser.upper.limits.B} \\
NGC 2273 & 0.8 & 0.06 &  160 & 0.5 & \citet{pesce:2015.agn.maser.upper.limits.B} \\
NGC 3393 & 3.1 & 0.04 &  300 & 0.5-1.5 & \citet{pesce:2015.agn.maser.upper.limits.B} \\
%%%UGC 3789 & 1 & 0.04  & 300 & --- & \citet{pesce:2015.agn.maser.upper.limits.B} \\ (only radial, don't know where from)
NGC 6323 & 0.9 & 0.1 & 300 & 0.2 &  \citet{pesce:2015.agn.maser.upper.limits.B} \\
NGC 2960 & 1.2 & 0.05 &  720 & 0.35 & \citet{pesce:2015.agn.maser.upper.limits.B} \\ %  (Mrk 1419)
ESO 558-G009 & 1.8 & $>$0.007  & 310 & 0.7 & \citet{pesce:2015.agn.maser.upper.limits.B} \\
Circinus & 0.17 & 0.2 & 150-360 & 0.1-0.4 & \citet{mccallum:2007.circinus.maser.B.limits} \\
\hline\hline}
\end{footnotesize}

\begin{figure}
	\centering
	\includegraphics[width=0.99\columnwidth]{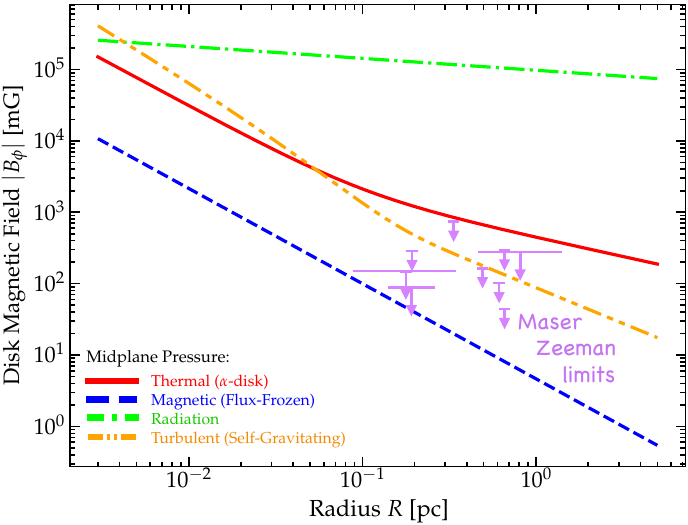} 
	\caption{Corresponding predictions as (Fig.~\ref{fig:vc}) for the midplane in-plane magnetic field strengths $\langle | B_{\phi}|\rangle$ (assuming a Maxwell stress comparable to Reynolds/accretion stress for each model). We overplot upper limits from maser Zeeman constraints (Table~\ref{tbl:Bfields}). Magnetically-dominated disks actually predict the weakest absolute $|{\bf B}|$ of any of the models, owing to their much lower overall densities (Fig.~\ref{fig:HR.rho}). This appears to be clearly favored by the maser constraints at $\gtrsim 0.1\,$pc.
	\label{fig:B}}
\end{figure}

\subsubsection{Constraints from Zeeman Observations on Magnetic Field Strengths}
\label{sec:thermal:mag}

In the ``arbitrary'' thermal-pressure dominated models, we have no prior for the magnetic field strengths. But of course in the classic, physically-motivated and widely-used SS73 or $\alpha$-disk models, the stress $\alpha \ne 0$ directly reflects the magnetic stresses in the disk (generally believed to arise from e.g.\ the magnetorotational instability [MRI] in the disk), with $\alpha \sim |\langle B_{\phi} B_{R} \rangle| /(4\pi \rho c_{s}^{2}) \lesssim  \langle B_{\phi}^{2} \rangle/(4\pi \rho c_{s}^{2})$ \citep{shakurasunyaev73,balbus.hawley.review.1998,bablus:mhd.angular.momentum.transport.review} (note this also includes contributions from Reynolds stresses, but these are fixed in ratio and modestly sub-dominant in saturated MRI turbulence; \citealt{pessah:2006.mri.signature.ratio.maxwell.reynolds}). So we can estimate a minimum toroidal magnetic field strength predicted by these models (as did SS73), giving: 
\begin{align}
\nonumber B_{\phi}^{\rm therm,\,pred} \gtrsim 1.5 {\rm G} \times {\rm MAX} {\Bigl[} 
& \frac{m_{7}^{69/80} \dot{m}^{17/40} \alpha_{0.1}^{1/20}}{r_{0.1}^{21/16}} 
\ , \ \\
\label{eqn:B.therm} & \frac{m_{7}^{23/32} \dot{m}^{23/32}}{r_{0.1}^{17/32} \alpha_{0.1}^{3/8}} 
{\Bigr]}
\end{align}

This is quite large, and potentially in tension (though much less dramatically so than the disk mass constraints above) with observational constraints from Zeeman splitting in masers, summarized in Table~\ref{tbl:Bfields} and shown in Fig.~\ref{fig:B}. We focus on direct maser Zeeman detections or upper limits here, since these provide a more or less model-independent upper limit to $B_{\phi}$ most robustly, as compared to the methods in \citet{silantev:2010.alt.B.estimation.method} and \citet{piotrovich:2021.B.estimation.from.disk.modeling.assuming.ss73} for inferring ``disk magnetic field strengths'' which (1) are sensitive to measurements coming from radii near-horizon (near the jet-launching region) and (2) are highly model dependent (e.g.\ \textit{assuming} SS73 to infer $|{\bf B}|$ from other properties like the accretion rates and spectral shape). Restricting to the direct maser constraints, we find specifically for (NGC 3079, 4258, 1194, 2273, 3393, 6323, 2960, ESO 558-G009, Circinus) the predicted $|B_{\|}^{\rm therm}|$ are $\sim (33-110,\,30-110,\,140,\,72,\,80-150,\,210,\,100,\,24,\,64-170)\,{\rm mG}$ from Eq.~\ref{eqn:B.therm}. One case, NGC 3079, is marginally in tension: the \citet{vlemmings:2007.maser.b.limits} upper limit is a $3\sigma$ bound and corresponds to the lower end of the predicted range for the lowest accretion rates allowed for this object. Even taken at face value this is only a $2-3\sigma$ tension, and is nowhere near as unambiguous as the mass-based comparisons above.

\subsection{Magnetic-Pressure Dominated Disks (Hyper-Magnetized, Flux-Frozen Disks)}
\label{sec:magnetic}

It is worth explicitly comparing the present picture to the classic analysis of \citet{goodman:qso.disk.selfgrav}, who identified the same outer-disk self-gravity problem highlighted here and enumerated a short list of candidate solutions. Goodman considered magnetic torques (``from a wind or corona''), gravitational torques from bars or global spirals, and star-formation-regulated support, and argued the first could ``increase the accretion speed and reduce the density of the disc'' but then discarded the magnetic solution on the grounds that ``the accretion speed is at most sonic, so that instability still sets in beyond about a parsec.'' We stress that this sonic-speed limit is not fundamental: it is an implicit consequence of assuming the disk cannot sustain plasma $\beta \ll 1$ throughout, so that the Alfv\'en speed limiting the accretion velocity is at most comparable to $c_{s}$. The hyper-magnetized/flux-frozen regimes proposed in \citet{begelman.pringle:2007.acc.disks.strong.toroidal.fields,johansen.levin:2008.high.mdot.magnetized.disks,gaburov:2012.public.moving.mesh.code,hopkins:superzoom.overview,hopkins:superzoom.disk,hopkins:superzoom.analytic} explicitly violate that assumption: they are trans-Alfv\'enic and super-sonic, with $t_{\rm acc} \sim 10\,\Omega^{-1}$ at all radii from the BHROI to near-horizon. Recent simulations \citep{hopkins:superzoom.overview,hopkins:superzoom.disk,guo:2024.fluxfrozen.disks.lowmdot.ellipticals,kaaz:2024.hamr.forged.fire.zoom.to.grmhd.magnetized.disks,squire:2024.mri.shearing.box.strongly.magnetized.different.beta.states,shi:2024.imbh.growth.feedback.survey,wang:2025.hypermagnetized.circumbinary.disk.flux.frozen.cavity.to.pc.scales,luo:2024.magnetically.dominated.disk.like.our.zoomins.zoomin.on.first.supermassive.star.situation,guo:2025.idealized.sphere.collapse.sims.hypermagnetized.disks.resolution.dependent.on.resolving.thermal.scale.height.but.limited.physics} show this regime is self-consistently achievable when the outer boundary (the larger-scale ISM) supplies magnetic flux. So we agree with and expand upon \citet{goodman:qso.disk.selfgrav} regarding the problems of thermal disk models; the difference is that in the intervening two decades there has been considerable development of the theory and simulation of strongly-magnetized disks.

\subsubsection{Agreement with Observed Dynamical Constraints on Disk Mass Profiles}
\label{sec:magnetic:mass}

What if instead the disk were magnetic pressure dominated, so $P_{\rm tot} \approx P_{\rm mag} = B^{2}/8\pi \sim (1/2)\,\rho\,V_{A}^{2}$? For disks in this regime, the self-consistent model for structure assuming $P_{\rm tot} \approx P_{\rm mag}$ with magnetic field strengths determined by flux-freezing is presented in \citet{hopkins:superzoom.analytic}. This gives the prediction
\begin{align}
\label{eqn:mgas.mag} \frac{M_{\rm gas}^{\rm mag}(<R)}{M_{\rm BH}} \sim 10^{-3}\frac{\dot{m} r_{0.1}^{7/6} r_{ff,\,5}^{1/3}}{\alpha_{0.1}^{2} m_{7}^{1/2}} \sim 10^{-5} \frac{\dot{m} r_{0.1}^{7/6}}{\alpha^{2} m_{7}^{1/3}}
\end{align}
(here our parameter $\alpha$ serves the same role as the parameter $\psi$ in \citealt{hopkins:superzoom.analytic}). This is orders-of-magnitude smaller than $M_{\rm gas}^{\rm therm}$, because the Maxwell stresses and turbulence are much stronger than an SS73 disk which is limited by $P_{\rm mag} \sim P_{\rm turb} \ll P_{\rm therm}$. Obviously this is easily allowed by present constraints on $\Delta$, for any reasonable $\alpha$, and we see this plainly in Figs.~\ref{fig:vc.obs}-\ref{fig:HR.rho}.

Even considering the couple of systems for which a claimed disk mass is detected from \S~\ref{sec:thermal:default}, the magnetized models appear more consistent with the observations. For Circinus, recall the direct imaging of neutral gas mass gives $M^{\rm Circ}_{\rm disk} \sim 3500-6100\,{\rm M}_{\odot}$ at $r<0.27\,$pc \citep{izumi:2023.imaging.nuclear.gas.disk.circinus.accretion.rate}. The prediction from Eq.~\ref{eqn:mgas.mag} for the same BH mass and accretion rate is $M_{\rm gas}^{\rm mag} \sim 3000\,\alpha_{0.1}^{-2} {\rm M}_{\odot}$, in remarkably good agreement for a Maxwell stress approximately equal to $\sim 10\%$ of the total magnetic pressure (very similar to that predicted in the simulations in \citealt{hopkins:superzoom.disk}). The more extreme, albeit indirect claims from kinematics for NGC 1068 and NGC 3079 \citep{lodato:2003.ngc.1068.agn.massive.acc.disk.some.evidence.rules.out.ss73,kondratko:2005.3079.selfgrav.disk.mass.masers} are larger than the predicted disk mass from Eq.~\ref{eqn:mgas.mag} by factors of $10-100$. However, as noted by those authors themselves as well as \citet{kuo:2018.maser.bh.masses.much.closer.to.keplerian.than.some.have.argued.percent.deviations.only}, there are alternative explanations, for example that the deviation from a Keplerian rotation curve is caused by stellar mass becoming significant interior to $\sim 1\,$pc in these systems, or the known irregular kinematics of the jet/wind system in 3079. In these cases the claimed disk masses should be treated as upper limits, making the prediction of Eq.~\ref{eqn:mgas.mag} consistent with the observations. And indeed \citet{gallimore:2023.ngc.1068.no.massive.acc.disk.just.eccentric.warp} revisited 1068 with higher-resolution and more extensive datasets and argued $M^{1068}_{\rm disk} < 10^{5}\,{\rm M}_{\odot}$, completely consistent with Eq.~\ref{eqn:mgas.mag} but smaller by a factor of $\sim 1000$ than the predictions for any thermal-pressure dominated disk (\S~\ref{sec:thermal:default}). They note that the apparent deviations from Keplerian rotation are more consistent with a coherent eccentric mode with a few-percent amplitude, exactly the sorts of modes that appear to be ubiquitous in magnetically-dominated disks \citep[see the examples in][]{hopkins:superzoom.overview,hopkins:superzoom.disk}. 

Briefly, we note that sometimes it has been argued that spiral or eccentric structure in an accretion disk implies significant disk self-gravity \citep{maoz:1995.spiral.structure.4258.implies.disk.mass.if.selfgrav,gallimore:2023.ngc.1068.no.massive.acc.disk.just.eccentric.warp}, but this is incorrect. It has been known for decades that even in the test-particle limit of \textit{zero} self-gravity in a perfectly Keplerian potential, strong spiral and/or eccetric modes can easily be powered \textit{directly} by the MRI \citep{heinemann.papaloizou:2009.spiral.mode} or by other non-axisymmetric magnetic modes in non-self-gravitating but strongly magnetized disks \citep{das:2018.pessah.psaltis.limit.mri}, or seeded by random fluctuations in disks with strong magnetic fields \citep[see references in][]{hopkins:superzoom.overview,hopkins:superzoom.disk}, or can propagate in from $R=\infty$ (much larger radii where self-gravity may not be negligible) without any barrier \citep{adams89:eccentric.instab.in.keplerian.disks,tremaine:slow.keplerian.modes,jacobs:longlived.lopsided.disk.modes,touma:keplerian.instabilities,bacon:m31.disk,hopkins:inflow.analytics,hopkins:slow.modes,hopkins:cusp.slopes}. And indeed such modes are ubiquitously seen in simulations with non-linearly large amplitudes even when self-gravity is disabled in the simulations \citep[see][]{hopkins:zoom.sims,hopkins:m31.disk,hopkins:qso.stellar.fb.together,gaburov:2012.public.moving.mesh.code,jiang:2019.superedd.sims.smbh.prad.pmag.modest.outflows,kudoh:2020.strong.b.field.agn.acc.disk.sims.compare,davis:2020.mhd.sim.acc.disk.review,kaaz:2022.grmhd.sims.misaligned.acc.disks.spin}. So existence of even non-linearly high-amplitude spiral structure or eccentric orbits does not imply any meaningful constraint, in and of itself, on the mass of the accretion disk.

From the same magnetically-dominated accretion disk models in \citet{hopkins:superzoom.analytic}, we immediately obtain the related predictions for the mean density $n \sim 10^{6}\,m_{7}^{3/4}\,\dot{m}\,r_{0.1}^{-2}\,\alpha^{-3}$, but note that the disk is necessarily supersonically turbulent and multiphase so the dense gas has densities between this and $n\,(V_{\rm turb}/c_{s})^{2}$ or $n \sim (10^{6}-10^{10})\,\alpha^{-3}\,{\rm cm^{-3}}$ (see \citealt{hopkins:superzoom.overview,hopkins:superzoom.disk,hopkins:multiphase.mag.dom.disks}). The model also predicts $\tau \lesssim 1$ for both scattering and absorption of the maser and BLR emission at these wavelengths at all radii $\gg$\,light-days, and at maser radii $\sim 0.1\,$pc temperatures $T \sim T_{\rm eff} \sim 200-1000$\,K at $\sim 0.1$\,pc \citep{hopkins:superzoom.overview,hopkins:multiphase.mag.dom.disks}, depending on whether we consider the colder more-shielded midplane or partially-illuminated surface layers. And the disk is thick and flared, with $H/R \sim \alpha m_{7}^{-1/12}$ \citep{hopkins:superzoom.analytic,hopkins:superzoom.disk}. More detailed comparisons are in \citet{hopkins:superzoom.overview,hopkins:superzoom.disk} and Bardati et al.\ (in prep.), but at the order-of-magnitude level, the magnetically-dominated model is remarkably consistent with both the properties of masing gas at $\sim 0.1-1\,$pc and BLR emitting gas at $\sim 1-100\,$ld ($\sim 0.001-0.1$\,pc).

\subsubsection{Effects of Pressure Support at Large $H/R$}
\label{sec:magnetic:pressurefx}

For these magnetized disks, $H/R$ is sufficiently large ($\gtrsim 0.1$) that one might worry about pressure support causing non-Keplerian motion, even in the test particle limit (i.e.\ the asymmetric drift correction being large). Generically, for a magnetically-supported disk, this gives:
\begin{align}
\label{eqn:nonthermal.support.Vrot.B} V_{\rm rot} &\approx V_{c} \sqrt{1 + \frac{1}{\rho V_{c}^{2}} \frac{\partial {P_{\rm eff}}}{ \partial \ln{R}} } \approx V_{K} \left[1 + \frac{h^{2}}{2} \left( 1 + \frac{\partial \ln B_{\phi}}{\partial \ln R} \right) \right]
\end{align}
where $h=H/R \approx V_A/2\,V_{K}$. Importantly, if $H/R \sim $\,constant \textit{or} $\partial \ln B_{\phi}/\partial \ln R \approx -1$, there  is \textit{no} measurable correction: the shape of the rotation curve is exactly Keplerian at all $r$ (just with a potentially slightly different normalization, which is degenerate with a slightly different BH mass). But in the analytic models in \citet{hopkins:superzoom.analytic} and simulations in \citet{hopkins:superzoom.overview,hopkins:superzoom.disk}, both of these conditions are approximately true: the favored models give $H/R$ varying between constant ($\propto R^{0}$) and $\propto R^{1/6}$, and $-4/3 \le \partial \ln B_{\phi}/\partial \ln R \le -1$. So in-practice calculating the expected non-Keplerian motions from pressure support from \textit{any} of the analytic family of models in \citet{hopkins:superzoom.analytic} or the simulations in \citet{hopkins:superzoom.overview,hopkins:superzoom.disk} gives deviations from $V_{K}$ below percent-level. 

Moreover, even those (very weak) effects are quite distinct from the upper limits to increasing $V_{\rm c}$ we are focused on here. Because the disk can flare ($H/R$ increases with $R$ and $\partial \ln B_{\phi}/\partial \ln R \le -1$), this predicts a very slightly \textit{steeper} decline of $V_{\rm rot}$ with $R$ than Keplerian, meaning that the sense of the deviations predicted is actually \textit{more} consistent with the upper limits in the literature. In practice this deviation is basically undetectable: over a 1\,dex range from $\sim 0.1-1\,$pc for a $\sim 10^{7}\,M_{\odot}$ BH, the favored model from \citet{hopkins:superzoom.analytic} with the maximal deviation ($\partial \ln B_{\phi}/\partial \ln R=-4/3$) gives a rotation curve which would be formally best-fit by $V_{\rm rot} \propto r^{-0.505}$ instead of $r^{-0.5}$, well within observational uncertainties. We can also confirm this directly in the full global simulations of \citet{hopkins:superzoom.disk}. There, the deviations predicted from the magnetic pressure support are similarly un-measurably small (not surprising, since the analytic scaling of the favored model in \citealt{hopkins:superzoom.analytic} is a good fit to the simulations in \citealt{hopkins:superzoom.disk}), smaller than this upper limit (since $H/R$ is slightly closer to constant, see Fig.~5 in \citealt{hopkins:superzoom.disk}). There there are much larger deviations at these radii from Keplerian driven by eccentric motions in the disk: evidence for these is also commonly seen in maser disks (see discussion above), but of course this has nothing to do with the mass profile and enters differently from the disk mass limits above (in how it modifies the apparent rotation curve, since it causes a global asymmetry) and therefore does not change any of our conclusions here.

Note the $ 1 + {\partial \ln B_{\phi}}/{\partial \ln R} $ term in Eq.~\ref{eqn:nonthermal.support.Vrot.B} and vanishing of the corrections to Keplerian motion for $B_{\phi} \propto R^{-1}$ arises because of the competing magnetic tension and pressure terms in $B_{\phi}$ in the radial momentum equation. Such cancellation is not possible for e.g.\ a thermal or radiation-pressure dominated disk, which would predict much larger deviations from Keplerian orbits if $H/R$ were large. So again, the nearly-Keplerian behavior observed further favors magnetically-dominated disks.

\subsubsection{Measurement of Magnetic Field Strengths from Maser Zeeman Splitting}
\label{sec:magnetic:Bconstraints}

As in \S~\ref{sec:thermal:mag}, we can also attempt to compare the predicted in-plane magnetic field strengths from these models directly to the maser Zeeman upper limits compiled in Table~\ref{tbl:Bfields}. From the same models giving Eq.~\ref{eqn:mgas.mag} \citep{hopkins:superzoom.analytic}, we have the predicted: 
\begin{align}
\label{eqn:B.mag} B_{\phi}^{\rm mag,\,pred} &\sim 0.1 {\rm G} \, \frac{m_{7}^{19/24}\dot{m}^{1/2}} {\alpha^{1/2} r_{0.1}^{4/3}}
\end{align}
For (NGC 3079, 4258, 1194, 2273, 3393, 6323, 2960, ESO 558-G009, Circinus) the predicted $|B_{\|}^{\rm mag}|$ are $\sim 
(0.7-1.7,\,0.8-3.4,\,5.7,\,2.4,\,1.3-5.7,\,12,\,4.9,\,1.0,\,1.7-11)\alpha^{-1/2}\,{\rm mG}$, all well below the observed upper limits even for rather low $\alpha$. 
We can also compare to the \citet{lopezrodriguez:2013.torus.upper.limits.B} estimate of $|{\bf B}|$ from polarization on ``torus'' scales in IC 5063 ($m_{7} \sim 28$; $\dot{m}\sim 0.02-0.1$), with $|B_{\|}| \sim 12-130\,$mG inferred from polarization of an unknown hot dust emitting region at $<250\,$pc, probably associated with the dust sublimation radius at $\sim 1\,$pc as argued by  \citet{lopezrodriguez:2013.torus.upper.limits.B}, where Eq.~\ref{eqn:B.mag} would predict $|B_{\|}^{\rm mag}| \sim (9.2-20)\,\alpha^{-1/2}\,{\rm mG}$ (for the uncertain range of $\dot{m}$ quoted above), reasonably consistent for a plausible range of $\alpha$.

From comparison of Eq.~\ref{eqn:B.therm} \&\ \ref{eqn:B.mag}, more explicitly in Fig.~\ref{fig:B}, we can immediately see the important point emphasized in \citet{hopkins:superzoom.analytic} and \citet{hopkins:superzoom.disk}. Namely, that the in-plane $B_{\phi}$ or $B_{\|}$ is actually \textit{smaller}, for the same BH mass and accretion rate, in the magnetically-dominated models as compared to the thermal-pressure-dominated models. This occurs because the thermal-dominated models are so much thinner and higher-surface density (owing to low temperatures making the upper limit to the stress $\Pi$ very small, hence requiring a large mass to support even a modest accretion rate), that their physical 3D midplane densities $\rho$ are many orders of magnitude larger than those predicted by the magnetically-dominated disk models, and so maintaining even a small \Alf\ speed or Maxwell stress in those thermal-dominated models requires an order-of-magnitude larger absolute value of $|{\bf B}|$.

\subsection{Radiation-Pressure Dominated Disks}
\label{sec:radiation}

Now consider a radiation-pressure dominated disk, $P_{\rm tot} \approx P_{\rm rad}$. The standard radiation-pressure dominated solution requires the disk be optically-thick, but it also becomes strongly turbulent which removes strong stratification, giving $P_{\rm rad} \sim (4\sigma_{B}/3 c)\,T_{\rm rad}^{4} \sim (4\sigma_{B}/3 c)\,T_{\rm eff}^{4}$. The standard ``self-consistent'' solutions given in either \citet{shakurasunyaev73} or \citet{abramowicz:1988.slim.disks} give a similar predicted scaling: 
\begin{align}
\frac{M_{\rm gas}^{\rm rad}(<R)}{M_{\rm BH}} \sim 400 \frac{r_{0.1}^{7/3}}{m_{7}^{5/3} \dot{m}^{2/3} \alpha_{0.1}^{2/3}}
\end{align}
with $V_{c} \propto r^{+2/3}$, which can be immediately ruled out in Figs.~\ref{fig:vc.obs}-\ref{fig:HR.rho}.\footnote{At much smaller radii where the disk mass eventually becomes smaller than the BH mass, this model predicts $M_{\rm gas}/M_{\rm BH} \sim 5300 r_{0.1}^{7/2}\,m_{7}^{-5/2}\,(\dot{m}\alpha_{0.1})^{-1}$, but that would only apply at radii interior to $r \ll 0.007 {\rm pc} \, (\alpha_{0.1} \dot{m})^{2/7} m_{7}^{5/7}$, well interior to the radii of interest here, and moreover even then this scaling would still be ruled out by $\Delta_{0.01} \sim 1$ (nearly-Keplerian rotation), down to even smaller radii $r \lesssim {\rm 1} (\alpha_{0.1} \dot{m})^{2/7} m_{7}^{5/7}$\,light-day, interior to the BLR.}

We note, as for the thermal case in \S~\ref{sec:thermal}, that the asymmetric-drift correction to $V_{c}$ for this model is $\delta V_{c} / V_{c} \sim (H/R)^{2} \lesssim 10^{-4}$, negligible on the scale of Fig.~\ref{fig:vc.obs}.

More generally, a radiation-pressure supported disk, by definition, $\kappa\,F/c = g \approx \Omega^{2}\,H$, so if we define the luminosity $L(R) \approx 2\pi\,R^{2} F$ as above and the usual $L_{\rm Edd}=4\pi G M_{\rm BH} m_{p} c /\sigma_{T}$, our constraint becomes: 
\begin{align}
\frac{L(R)}{L_{\rm Edd}} \gtrsim \frac{H/R}{\kappa/\kappa_{\rm es}} \gg 0.1\frac{\dot{m}^{1/2}\,r_{0.1}^{3/4}}{\alpha_{0.1}^{1/2}\,\Delta_{0.01}^{1/2}\,m_{7}^{1/4}}\,\frac{\kappa_{\rm es}}{\kappa} 
\end{align}
Thus unless $\kappa \gg \kappa_{\rm es}$, it is not possible to hold up the disk and obey the observed constraints without producing far more radiation than observed, with \textit{most} of the radiation coming from large radii, and being vastly colder than observed (characteristic radiation temperatures $\lesssim 1000\,$K). This is clearly ruled-out observationally but is also unphysical, since there is nothing that could power such large luminosities at large radii (it is many orders-of-magnitude larger than the accretion luminosity). 

Even if one invokes large dust opacity at the very largest radii $\gtrsim$\,pc, it would not in general bring $L(R)$ required here below observational limits. But for many inner maser regions, the dust is clearly sublimated (there is nowhere near as much dust observed as there would be if it were not). If the dust is sublimated and the gas atomic/molecular with observed maser temperatures $\ll 10^{4}\,$K, then the maximum opacity is $\kappa \sim \kappa_{\rm mol} \sim 0.001\,(Z/Z_{\odot})$, so we have $L(R) \gtrsim 100\,L_{\rm Edd}\,r_{0.1}^{3/4}$ -- i.e.\ the observed luminosities immediately rule out a radiation-pressure supported disk. More rigorously, since the masing radii cannot be optically-thick to self-absorption (or they would not be observed as they are), we can set a strong upper limit on their opacity at their radii, giving $\kappa/\kappa_{\rm es} \ll 0.02/[(H/R)\,n_{8}\,r_{0.1}\,\mu_{\rm mol}]$ for a masing gas density $n=n_{8} 10^{8}\,{\rm cm^{-3}}$, which means we \textit{must} at a maser radii have 
\begin{align}
\frac{L(R)}{L_{\rm Edd}} \gg 0.5 \frac{\dot{m} n_{8} R_{0.1}^{5/2}} {m_{7}^{1/2} \alpha_{0.1} \Delta_{0.01} }
\end{align}
in order to have a radiation-pressure supported disk. Again, this is immediately ruled-out.

\begin{footnotesize}
\ctable[caption={{\normalsize Summary of Predictions of Different Accretion Disk Pressure Sources}.\label{tbl:summary}},center,star]{lccccl}{
\tnote[ ]{For each row, we assume the pressure holding up the accretion disk at large radii ($\gg 0.01\,$pc) comes primarily from the given source. 
Columns: {\bf (1)} Pressure source. {\bf (2)} Shape of the rotation curve at these radii. {\bf (3)} Other immediate contradictions with observations this would predict, if we forced the parameters (e.g.\ $T$, $L$, etc.) to have values consistent with the maser kinematics. {\bf (4)} Theoretical contradictions if one attempts to support the disk with such pressure.}}{
\hline
Dominant Pressure & Slope $V_{c} \propto R^{\zeta_{V}}$ & Observational Contradiction & Theoretical Contradiction \\
\hline
Radiation & $\zeta_{V} \sim +2/3$ (steeply-rising) & $L^{\rm outer\,disk} \gg L_{\rm Edd}$ always, most luminosity from $\gtrsim$\,pc & No physical radiation source \\
Thermal & $\zeta_{V} \sim +3/8$ (steeply-rising) & $L^{\rm outer\,disk} \gg L_{\rm Edd}$ always, $T_{\rm spec} \gg 10^{6}\,$K (spectra super-hot) & No physical heating source \\
Cosmic Ray & No consistent model & $\gamma$-ray luminosity $L_{\gamma} \gtrsim L_{\rm Edd}$ & No sufficient CR source \\
Gravitoturbulence & $-0.25 < \zeta_{V} < 0$ (flat or slow-falling) & Requires $Q \sim 100$, not $Q\sim 1$; no single $Q$ fits $H/R$ \&\ $M_{\rm disk}$ & What drives the turbulence?  \\
Magnetic & $-0.55 \lesssim \zeta_{V} \lesssim -0.5$ (near-Keplerian) & None & None \\
\hline\hline}
\end{footnotesize}

\subsection{Turbulent Ram-Pressure Dominated Disks}
\label{sec:turb}

\subsubsection{Allowed Strength of Turbulence and Conceivable Drivers in the Absence of Any Other Dominant Pressure}
\label{sec:turb:allowed}

In simulations and models of both magnetic and radiation-pressure dominated disks, the predicted turbulence is broadly near equipartition with these energy sources \citep{jiang:2019.superedd.sims.smbh.prad.pmag.modest.outflows,hopkins:superzoom.disk}. So, given that the magnetic pressure-dominated models appear completely consistent with observational constraints (\S~\ref{sec:magnetic}), it is perfectly plausible that turbulence could be an $\mathcal{O}(1)$ component of the ``required'' disk support. We can see this directly from Eq.~\ref{eqn:constraint} with $\Pi \sim \alpha P_{\rm tot} \sim \alpha P_{\rm turb} \sim \alpha \rho V_{\rm turb}^{2}$, the turbulent velocity must only exceed:
\begin{align}
\label{eqn:vturb.est} V_{\rm turb} \gtrsim 20\,{\rm km\,s^{-1}} \frac{\dot{m}^{1/2} m_{7}^{1/4} r_{0.1}^{1/4}}{\alpha^{1/2}\Delta_{0.01}^{1/2}}
\end{align}
This is much smaller than $V_{\rm K}$, and completely allowed by the more detailed kinematic and dynamical modeling of line dispersions/widths from maser and BLR imaging (not just the rotation curves in Figs.~\ref{fig:vc.obs}-\ref{fig:HR.rho}). 

So turbulent pressure being of order $P_{\rm tot}$, as a statement in and of itself, is allowed by the constraints in Figs.~\ref{fig:vc.obs}-\ref{fig:HR.rho}. But the question relevant for this section here is: ``Can the disk be supported \textit{entirely} by turbulence?''. In other words, can one have $P_{\rm tot} \sim P_{\rm turb}$ with no \textit{other} large pressure of the forms already reviewed, i.e.\ thermal, magnetic, radiation, cosmic ray, and other pressures all much smaller than turbulent? This means the turbulence would necessarily be highly supersonic and super-\Alf{ic}, which in turn means its dissipation time must be of order the crossing/turnover time $t_{\rm diss} \sim H / V_{\rm turb} \sim \Omega$ (if it is supporting the disk in the first place). So since it \textit{must} dissipate, by definition, on a timescale faster than the accretion time, it must also therefore be ``powered'' by something. But by definition in this regime the power source cannot be the thermal, magnetic, radiation, cosmic ray, neutrino, or chemical energy of the disk. That leaves only two energy sources which have been discussed in the literature: feedback from massive stars within the disk (e.g.\ SNe, radiation, stellar mass-loss) and gravity. Note that feedback from the central accretion disk would not vertically support the disk, and if it had the form of a wind, would change the rotation curves, and if it acted via radiation/cosmic ray acceleration/thermal heating would violate our requirement of $P_{\rm tot} \sim P_{\rm turb}$ alone. 

\subsubsection{Stellar Feedback-Driven Turbulence Is Ruled Out on These Scales}
\label{sec:turb:fb}

\begin{figure}
	\centering
	\includegraphics[width=0.99\columnwidth]{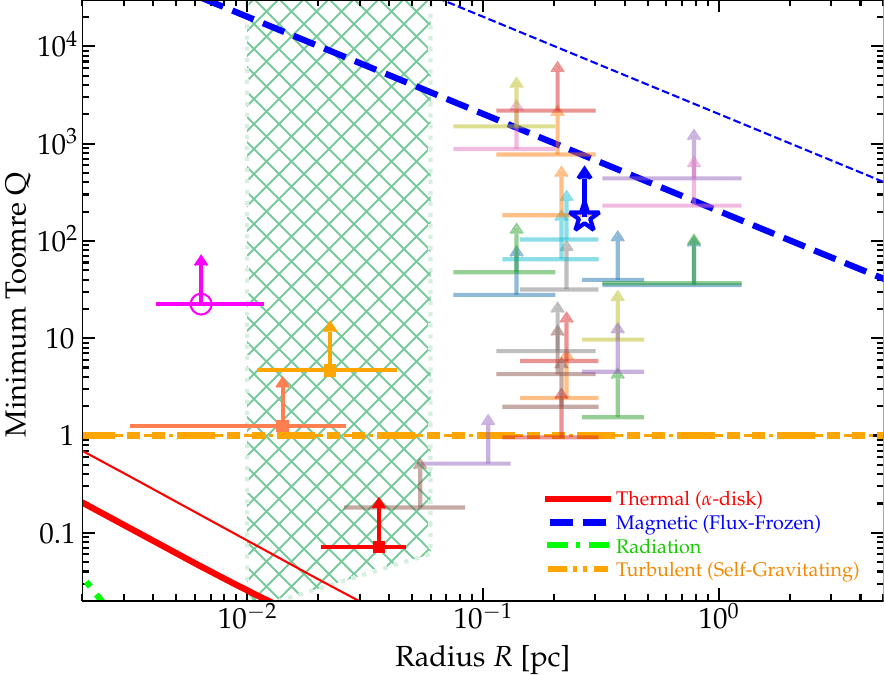} 
	\caption{Observational limits and predictions (as Fig.~\ref{fig:vc.obs}) from different models for the minimum Toomre $Q(r)$ at a given radius $r$ in the disk (\S~\ref{sec:turb:grav}), from kinematics alone (Eq.~\ref{eqn:Qmin}). We plot the ``turbulent'' line at $Q=Q_{0}\sim1$, the usual prediction for gravito-turbulent or self-regulating models. Below $\sim$\,pc scales, maser data and some BLR constraints require a minimum $Q \gtrsim 20-1000$, much larger than gravitoturbulent predictions, and many orders of magnitude larger than predicted for thermal or radiation-pressure supported disks, but consistent with magnetically-supported disk models.
	\label{fig:Q}}
\end{figure}

Feedback from stars at these radii is not viable for at least four reasons: 
(1) It is dynamically unstable when the dynamical time is $\ll 10^{6-7}\,{\rm yr}$ ($R \ll 100$\,pc), and cannot ``support'' a quasi-steady disk \citep{torrey.2016:fire.galactic.nuclei.star.formation.instability}. In brief, if one forms slightly ``too many'' massive stars (or they begin to blow away or exhaust or accrete even a small fraction of the disk gas so the disk gas mass decreases), the stars do not ``turn off'' until after their main-sequence lifetimes complete and they all finish exploding, so the feedback injection rate actually increases for this timescale (much longer than the dynamical time) and the blowout runs away.
(2) As many studies have shown \citep[][and references therein]{fall:2010.sf.eff.vs.surfacedensity,grudic:sfe.cluster.form.surface.density,grudic:max.surface.density,grudic:mond.accel.scale.from.stellar.fb,ma:2020.globular.form.highz.sims,hopkins:2021.bhs.bulges.from.sigma.sfr}, above a critical acceleration scale $|{\bf a}| \sim G\,M_{\rm bh}/R^{2} \sim 10^{-7}\,{\rm cm\,s^{-2}}$ ($R \lesssim 40 m_{7}^{1/2}\,{\rm pc}$), stellar feedback from standard stellar populations/evolution tracks cannot inject sufficient momentum to ``hold up'' the disk without the mass of stars greatly exceeding the total gas mass, thus star formation greatly exceeding accretion rates and depleting the disk (preventing accretion).
(3) Even if we ignore (1) and (2), a continuous feedback-regulated model predicts $Q\sim1$ \citep{thompson:rad.pressure,ostriker.shetty:2011.turb.disk.selfreg.ks,cacciato:2011.analytic.disk.instab.cosmo.evol,krumholz:2012.universal.sf.efficiency,hopkins:rad.pressure.sf.fb,hopkins:fb.ism.prop,orr:non.eqm.sf.model,orr:2020.resolved.dispersions.sfrs.correlations,orr:2021.fire.cmz.analog}, as discussed below for gravity, which Fig.~\ref{fig:Q} shows contradicts lower-limits from the maser kinematics.
(4) The implied massive stellar density and SNe rates in the disk (ignoring (1), (2), and (3), and imposing a steady-state model with the required parameters) would give stellar luminosities much larger than disk luminosities and stellar masses much larger than BH masses (i.e.\ $\Delta \gg 1$) in the outer disk, in addition to dense stellar ``cusps'' with orders-of-magnitude higher density than actually observed.\footnote{Specifically, to power the required turbulence would require, at any given time (assuming a standard IMF-integrated population; see \citealt{grudic:sfe.cluster.form.surface.density,shi:2022.hyper.eddington.no.bhfb}, and using the scalings assuming we are inside the BHROI), a young (zero-age main sequence) stellar mass interior to $r$ of 
$M_{\ast}^{\rm young}/M_{\rm BH} \gtrsim 10\,\dot{m} n_{8} r_{0.1}^{5/2} m_{7}^{-1/2} \alpha_{0.1}^{-1} \Delta_{0.01}^{-1} 
\gtrsim 110\,m_{7}^{3/4} \dot{m}^{1/2} \Delta_{0.01}^{1/2}\,r_{0.1}^{5/4} \alpha_{0.1}^{-1/2}$ (the latter using the minimum gas mass of the disk from \S~\ref{sec:theory} to estimate $n_{8}$), i.e.\ $\Delta \gg 1$, and a stellar luminosity 
$L_{\ast} \gtrsim 100\,L_{\rm BH}\,n_{8}\,r_{\rm pc}^{5/2}\,m_{7}^{-1}\,\alpha_{0.1}^{-1} \Delta_{0.01}^{-1}$. The former exceeds unity meaning there is no self-consistent $\Delta \ll 1$ (close-to-Keplerian) solution, nor is there a solution which features stellar luminosity interior to the BHROI less than AGN/disk luminosity. Integrating over the accretion history of the BH (using $\int \dot{M} dt = M_{\rm BH}$ by definition), this would imply a relic star cluster mass around the BH of at least 
$M_{\ast}^{\rm old}/M_{\rm BH} \gtrsim 30\,n_{8}\,r_{0.1}^{5/2}\,m_{7}^{-1/2}\,\alpha_{0.1}^{-1}\,\Delta_{0.01}^{-1}$, or $\Delta_{0.01} \gtrsim 50 \,n_{8}^{1/2}\,\alpha_{0.1}^{1/2} r_{0.1}^{5/4} m_{7}^{1/4}$, or stellar relic cusp surface density $\Sigma_{\ast} \gtrsim 10^{10}\,M_{\odot}\,{\rm pc}^{-2} m_{7}^{1/2} n_{8} r_{0.1}^{1/2} \alpha_{0.1}^{-1} \Delta_{0.01}^{-1}$ which is five orders-of-magnitude larger than the maximum observed $\Sigma_{\ast} \lesssim 10^{5}\,M_{\odot}\,{\rm pc^{-2}}$ in any circum-BH stellar cusp at these radii in the local Universe \citep{lauer:bimodal.profiles,cote:smooth.transition,jk:profiles,hopkins:cusps.ell,hopkins:maximum.surface.densities,grudic:max.surface.density}.}

\subsubsection{Standard Gravito-Turbulent Disks Are Not Consistent with Observations}
\label{sec:turb:grav}

So this leaves gravitational energy as the only remaining power source for a turbulent disk, if we do \textit{not} allow for something like a magnetically-dominated disk. This would be a ``gravitoturbulent'' or ``self-gravitating'' disk. The problem is that the most robust and universal prediction of such models is that if the disk is ``held up'' and accretion powered by Reynolds stresses from gravity-driven turbulence (\textit{without} another form of pressure comparable to or larger than the turbulence) then the disk must self-regulate at a turbulent Toomre $Q$ parameter near unity \citep[see references above and][]{paczynski:1978.selfgrav.disk,gammie:2001.cooling.in.keplerian.disks,kim.ostriker:2001.gravitoturb.galactic.disks.mhd.conditions,sirko:qso.seds.from.selfgrav.disks,thompson:rad.pressure,riols:2016.mhd.ppd.gravitoturb,forgan:2017.mhd.gravitoturb.sims,deng:2020.global.magnetized.protoplanetary.disk.sims.gravito.turb.leads.to.large.B.saturation.vs.mri}. In other words: 
\begin{align}
\label{eqn:Qpred.turb} Q_{\rm turb}^{\rm pred} \sim \frac{v^{\rm pred}_{\rm turb} \Omega}{\pi G \Sigma^{\rm pred}_{\rm gas}} = \frac{\Omega^{2}}{2\pi G \rho^{\rm pred}} = Q_{0} \sim 1
\end{align}
Physically, if $Q \gg 1$, then the disk self-gravity is negligible, and it will simply sit in a stable Keplerian orbit without powering any turbulence or accretion, until some cooling or other dissipation or energy transfer or new accretion reduces $Q$ to order-unity (while if $Q \ll 1$, these modes excite strong turbulence to bring $Q$ back to $\sim 1$). But Eq.~\ref{eqn:Qpred.turb} immediately predicts a gas density
\begin{align}
 n_{\rm turb}^{\rm pred} \sim 10^{11}Q_{0}^{-1} {\rm cm^{-3}} m_{7} r_{0.1}^{-3} \sim 10^{17}Q_{0}^{-1} {\rm cm^{-3}} m_{7} r_{\rm ld}^{-3},
\end{align}
 much higher than observed in the inner maser or broad line regions unless $Q_{0} \gg 1$.\footnote{And this is actually a \textit{lower} limit to $n_{\rm turb}^{\rm pred}$, as we assumed isotropic turbulence, while a more careful derivation should multiply it by a factor of $\sim \delta v_{R}^{\rm turb} / \delta v_{z}^{\rm turb}$, the radial-to-vertical turbulent velocity dispersion ratio, which is typically $\gg 1$ in gravito-turbulence (because the salient gravitational modes only drive in-plane motions $\delta v_{R,\,\phi}^{\rm turb}$; \citealt{hopkins:fb.ism.prop,hopkins:2013.accretion.doesnt.drive.turbulence,hopkins:superzoom.disk,orr:ks.law,orr:non.eqm.sf.model,su:2016.weak.mhd.cond.visc.turbdiff.fx}).}
More plainly, from Eqs.~\ref{eqn:rhomax} and \ref{eqn:Qpred.turb},  we see that a given measurement or upper limit to $\Delta$ implies a corresponding \textit{lower} limit to $Q$ in the disk: 
\begin{align}
\label{eqn:Qmin} Q & \gtrsim 20\,\frac{r_{0.1}^{3/4} \dot{m}^{1/2}}{m_{7}^{1/4} \alpha_{0.1}^{1/2} \Delta_{0.01}^{3/2}}\ , 
\end{align}
which ranges from $\sim 10-1000$ over the range of observations we consider at large radii, as shown in Fig.~\ref{fig:Q}.

Related, let us now specifically consider the self-consistent predictions for such models -- i.e.\ imposing not just $Q_{0}\sim$\,constant, but $\dot{M}\sim\,$constant with the only available stress, Reynolds stress $\approx \rho\,V_{\rm t}^{2}$ driving accretion. This predicts $V_{\rm turb} \sim (3\,Q_{0} G \dot{M}_{\rm BH}/2)^{1/3}$ and 
\begin{align}
\nonumber \frac{V_{c,\,{\rm self}}^{\rm turb}}{V_{K}} &\sim 0.3 \dot{m}^{1/6} m_{7}^{-1/6} Q_{0}^{-1/3} r_{0.1}^{1/4} \\ 
\label{eqn:mdisk.turb} \frac{M_{\rm disk}^{\rm turb}}{M_{\rm BH}} &\sim 0.1 \dot{m}^{1/3} r_{0.1}^{1/2} m_{7}^{-1/6} Q_{0}^{-2/3} \\
\label{eqn:HR.turb} \frac{H_{\rm turb}^{\rm pred}}{R} &\sim 0.016 \frac{Q_{0}^{1/3} r_{0.1}^{1/2}}{m_{7}^{1/6}}\ ,
\end{align}
a much larger deviation from Keplerian rotation, and much thinner $H/R$, than allowed by maser observations (Eqs.~\ref{eqn:constraint}-\ref{eqn:HRmin}) for $Q_{0} \sim 1$. So the densities, disk masses, and scale-heights observed in maser data require that these models must self-regulate to $Q_{0} \gtrsim 100$ (not $Q_{0} \sim 1$) to be consistent with observations at radii $\sim 0.1-1\,$pc. 
We stress that the {\em absolute} $V_{c}$ in this regime is {\em flat or weakly falling} with radius -- not rising -- consistent with the slope tabulated in Table~\ref{tbl:summary} ($-0.25 \lesssim \zeta_V \le 0$). The ratio $V_{c}/V_{K}$ in the above expression rises as $r^{1/4}$ because $V_{K}$ falls faster than $V_{c}$, not because $V_{c}$ itself increases. In the fully self-gravity-dominated limit at large $R$, $V_{c}$ flattens entirely (as in \citealt{kondratko:2005.3079.selfgrav.disk.mass.masers} and references therein). This is distinct from the thermal ($\zeta_V = +3/8$) and radiation ($\zeta_V = +2/3$) cases, where $V_{c}$ itself truly rises with $R$.

Similarly, for the direct estimate in Circinus from \citet{izumi:2023.imaging.nuclear.gas.disk.circinus.accretion.rate} of $M^{\rm Circ}_{\rm disk} \sim 3500-6100\,{\rm M}_{\odot}$ at $r<0.27\,$pc, we have $M_{\rm disk}^{\rm turb} \sim 2 \times 10^{5}\,{\rm M}_{\odot} \,Q_{0}^{-2/3}$ predicted, requiring $Q_{0} \sim  220 - 500$ at these radii to fit the observations.\footnote{The scalings here and in \S~\ref{sec:turb:mag} assume $Q_{0} \gtrsim 1$ which gives $M_{\rm disk} < M_{\rm BH}$ at the radii of interest. For $Q_{0}\ll1$, where at large radii $M_{\rm disk} \gg M_{\rm BH}$, these scalings are modified to $M_{\rm disk}^{\rm turb} = 4\,3^{2/3} \dot{M} R\,(G \dot{M})^{-1/3} Q_{0}^{-4/3} \sim 0.002 \dot{m}^{2/3} r_{0.1} m_{7}^{-1/3} Q_{0}^{-4/3}$, $n^{\rm turb} \sim 2\times 10^{8}\,{\rm cm^{-3}} (m_{7} \dot{m})^{2/3} r_{0.1}^{-2} Q_{0}^{-7/3}$, $H^{\rm turb}/R \sim 0.33\,Q_{0}$, $B_{\phi}^{\rm turb} \sim 0.06\,Q_{0}^{-5/6} (E_{\rm B}/E_{\rm turb})^{1/2} (m_{7}\dot{m})^{2/3} r_{0.1}^{-1}$.}

Even the systems which have claims for apparent self-gravitating disks (NGC 1068 and 3079) do not appear to fit the self-consistent gravito-turbulent models. Consider NGC 3079: if we take the implied disk properties from \citet{kondratko:2005.3079.selfgrav.disk.mass.masers} at face value (interpreting the deviations from Keplerian rotation as indicating the disk mass), then fitting the apparently observed disk mass with Eq.~\ref{eqn:mdisk.turb} would require $Q_{0} \sim 0.04$,\footnote{When $M_{\rm disk}^{\rm turb} \gtrsim M_{\rm BH}$, Eq.~\ref{eqn:mdisk.turb} is modified to $M_{\rm disk}^{\rm turb} = 4\,3^{2/3} \dot{M} R\,(G \dot{M})^{-1/3} Q_{0}^{-4/3}$, but this requires $Q_{0} \lesssim 0.008\,\dot{m}^{1/2} r_{0.1}^{3/4} m_{7}^{1/4}$.} (similar to the estimates of $Q_{0}$ from the gas densities in \citet{kondratko:2005.3079.selfgrav.disk.mass.masers}, using Eq.~\ref{eqn:Qpred.turb}, which require $Q_{0} \sim 0.01$). But at the same time, the scale-height constraints in \citet{kondratko:2005.3079.selfgrav.disk.mass.masers} -- who note any such disk model must be very thick ($H/R \sim 0.15-1$) to fit their data -- require (Eq.~\ref{eqn:HR.turb}) $Q_{0} \sim 100-2000$. These disagree by $4-5$ orders of magnitude, indicating that the data cannot be fit by any constant-$Q_{0}$ turbulence-only model, and on top of this either value of $Q_{0}$ is orders-of-magnitude different from the prediction of gravito-turbulent or marginally self-gravitating disk models. The same problem appears in NGC 1068, with the disk-mass interpretation of \citet{lodato:2003.ngc.1068.agn.massive.acc.disk.some.evidence.rules.out.ss73} requiring $Q_{0} \sim 0.03$, while scale-height constraints from \citet{gallimore:2023.ngc.1068.no.massive.acc.disk.just.eccentric.warp} require $Q_{0} \gtrsim 100$.

On the other hand, in a scenario like the magnetically-dominated disks from \S~\ref{sec:magnetic:mass}, trans-\Alf{ic}, supersonic turbulence with velocities similar to Eq.~\ref{eqn:vturb.est} is not a problem with respect to these observations, as we have shown.

\subsubsection{Magnetic Field Strengths in Turbulence-Dominated Disks}
\label{sec:turb:mag}

Briefly, although the turbulence-dominated scenario assumes negligible magnetic pressure, in practice if we had a supersonic, super-\Alf{ic} turbulent disk, we would expect to amplify the magnetic fields to $\sim 10-50\%$ of equipartition with turbulence \citep[][and references therein]{federrath:supersonic.turb.dynamo,su:fire.feedback.alters.magnetic.amplification.morphology,guszejnov:fire.gmc.props.vs.z,martin.alvarez:2022.cosmological.turb.dynamo}. If we take the ``self-consistent'' turbulent models above and just treat this as a passive magnetic field, this predicts an in-plane typical magnetic field strength (assuming isotropic motions as the lower-limit to the in-plane component):
\begin{align}
B^{\rm turb,\,pred}_{\phi} &\gtrsim \frac{2.2}{\sqrt{3}} {\rm G} \left({\frac{E_{\rm B}}{E_{\rm turb}}}\right)^{1/2} \frac{m_{7}^{5/6} \dot{m}^{1/3}}{Q_{0}^{1/6} r_{0.1}^{3/2}}
\end{align}
Just like with thermal-pressure dominated disks in \S~\ref{sec:thermal:mag}, we see in Fig.~\ref{fig:B} the surprising result that (for plausible equipartition saturation strengths and $Q_{0} \sim 1$) this would predict \textit{larger} magnetic fields compared to the magnetically-dominated disks in \S~\ref{sec:magnetic:Bconstraints} (Eq.~\ref{eqn:B.mag})! The reason is the same: the model predicts (for $Q_{0} \sim 1$) much thinner, higher-density disks compared to the magnetically-dominated case, so to have trans-\Alf{ic} turbulence, the absolute value of $|B_{\|}|$ (in Gauss) must be correspondingly larger.

\subsection{Cosmic Ray, Neutrino, or Degeneracy-Pressure Dominated Disks}
\label{sec:cosmicray}

For completeness, consider some other sources of pressure which are already widely-agreed to be ruled out for AGN accretion disks but can support accretion disks in other astrophysical systems: cosmic rays (CRs), neutrinos, and degeneracy pressure. The latter (neutrinos and degeneracy pressure) are immediately ruled out by any $\Delta \ll 10^{10}$ or so, let alone $\ll 1$, because the densities required for degeneracy pressure and/or non-negligible neutrino opacity are so many orders of magnitude larger than any which are permitted by dynamical constraints here (and the emission would be wildly different as well). The case of CRs is slightly less obvious, and they can be in equipartition with other pressures in Solar-neighborhood ISM disks \citep{1998ApJ...506..329W,draine:ism.book,2018AdSpR..62.2731A}. In a CR-pressure dominated disk, $P_{\rm tot} \sim (1/3)\,u_{\rm cr}$ in terms of the total energy density of CRs $u_{\rm cr}$, dominated by $\sim 1-10\,$GeV protons. First note however that there are \textit{no} self-consistent solutions for a CR pressure-supported AGN accretion disk (this is why, of course, this is not usually discussed in this literature). If one equates the energy change from gravity or stresses with some $\Pi \le P_{\rm cr}$ to the CR loss rate, then the hadronic loss rate ($dE/dt d{\rm Vol} \approx 3\times10^{9} s^{-1} \,(\rho/{\rm g\,cm^{-3}}) \, P_{\rm cr}$) at any reasonable density would be vastly too large to be replenished (there is no steady-state solution except at extremely low densities), but at low densities where these would balance the CR loss time becomes dominated by diffusive escape with timescale $\sim H^{2}/\kappa$ (for a diffusion coefficient $\kappa$). But that, in turn, gives a solution for $H$ which is $\ll$ au at $r \sim $\,pc (i.e.\ vastly too thin) and is always (for \textit{any} $\kappa$ allowed in this regime) orders of magnitude smaller than the CR scattering/deflection length ($\sim 3\kappa/c$), so the CRs could not ``hold up'' the disk. But even if we ignored all of these arguments and simply used our lower limit to $\Pi/\rho$, we would require a CR energy density $u_{\rm cr} \sim 3\times10^{9}\,{\rm eV\,cm^{-3}} \dot{m} n_{8} m_{7}^{1/2} r_{0.1}^{1/2} \Delta_{0.01}^{-1}$, which is enormous compared to any reasonable estimates in galactic nuclei \citep{krumholz:2023.cosmic.ray.ionization.gamma.ray.loss.budgets}, and would imply an instantaneous $\gamma$-ray luminosity (using the hadronic scalings from \citealt{chan:2018.cosmicray.fire.gammaray}) at large radii of $L_{\gamma} \gtrsim 10^{46}\,{\rm erg\,s^{-1}}\,\dot{m}^{3/2}\,n_{8}^{2}\,m_{7}^{1/4}\,\alpha^{-1/2}\,\Delta_{0.01}^{-3/2}\,r_{\rm pc}^{17/4}$, far in excess of that observed. So we can safely rule out this class of models (as anticipated).

\subsection{External Potential (Stellar/ISM) Zone}
\label{sec:stellar}

At large enough radii from the SMBH (around the BH radius of influence [BHROI] $\sim G M_{\rm BH}/\sigma_{\ast}^{2}$ where $\sigma_{\ast}$ is the stellar nuclear velocity dispersion), stars will begin to dominate the matter density and total enclosed mass at those radii from the SMBH. The assumptions in \S~\ref{sec:theory} then no longer hold: of course, if the mass of stars becomes comparable to the BH itself, then $V_{\rm c}$ becomes non-Keplerian and $\Omega = V_{\rm c}/R$ includes the contribution from stars -- i.e.\ $\Delta \gg 1$. But more importantly, even if we somehow had $V_{\rm c} \approx V_{K}$, if stars dominate the \textit{local} density ($M_{\rm gas}(<R) \lesssim M_{\ast}(<R) \ll M_{\rm BH}$), then the stress driving accretion does not have to be an \textit{internal} stress in the gas. In other words, if we write $\dot{M} = 3\pi \nu_{\rm v,\,eff} \Sigma_{\rm gas}$ with $\nu_{\rm v,\,eff} \equiv \Pi_{\rm eff}/(\rho\Omega)$, then we can have $\Pi_{\rm eff} \rightarrow \Pi_{\rm ext} \gg P_{\rm tot,\,gas}$ in principle, where $\Pi_{\rm eff}$ represents an ``external stress'' from stars acting on gas. Indeed, in practice, in gas+stellar disks on ISM scales (where $M_{\ast} \gg M_{\rm gas}$), $\Pi_{\rm ext} \sim \eta_{\rm ext} \rho V_{c}^{2}$ (with $\eta_{\rm ext} \sim 0.01-1$; see \citealt{hopkins:zoom.sims,hopkins:inflow.analytics,daa:20.hyperrefinement.bh.growth}) is generally the dominant stress, coming primarily from two sources: (1) direct feedback effects (e.g.\ work done by expanding stellar winds or SNe on ambient gas), and (2) gravitational torques, i.e.\ non-axisymmetric torques from stars driving shocks and dissipation in the gas \citep{barneshernquist96,hopkins:disk.survival,hopkins:qso.stellar.fb.together,hopkins:superzoom.overview,hopkins:zoom.sims,cacciato:2011.analytic.disk.instab.cosmo.evol,querejeta:grav.torque.obs.m51,prieto:2016.zoomin.sims.to.fewpc.hydro.cosmo.highz,prieto:2017.zoomin.sims.agn.fueling.sne.fb,angles.alcazar:grav.torque.accretion.cosmo.sim.implications,daa:20.hyperrefinement.bh.growth,williamson:2022.gizmo.rhd.psph.sims.binary.smbh.torii.radiation.reduces.grav.torques,izquierdo:2023.massive.bh.galactic.nuclei.review.focus.on.grav.torques}. One then obtains $\dot{M} \sim \eta_{\rm ext} M_{\rm gas}(<R) \Omega$, so $M_{\rm gas}/M_{\rm BH} \sim \dot{m}/(\eta t_{S} \Omega)$ or $\sim 0.001 \eta_{\rm ext}^{-1} \dot{m} m_{7}^{1/4}$ at the BHROI. This then allows for efficient fueling of the SMBH even with a small disk \textit{gas} mass at these radii \citep{hopkins:zoom.sims,hopkins:inflow.analytics}. More importantly, for our purposes, it means that once the local density of stars becomes comparable to the density of gas and $\Delta \gtrsim 1$ (the behavior at/outside the BHROI), the system becomes ``ISM like'' rather than ``accretion disk like'' and the kinematic constraints above no longer translate to the same constraints regarding the nature of the accretion disk.

The literature on bar-driven and gravitational-torque-driven fueling of the inner parsec is extensive: \citet{shlosman:bars.within.bars} first proposed the ``bars-within-bars'' mechanism, \citet{shlosman:inefficient.viscosities} identified the outer-disk self-gravity problem and proposed fragmentation-driven reduction of the disk mass, \citet{begelman:direct.bh.collapse.w.turbulence} proposed that non-axisymmetric gaseous bars can drain the self-gravitating disk on a timescale $\sim 10\,\Omega^{-1}$ (closely analogous to the $t_{\rm acc}$ in the hyper-magnetized regime), and \citet{choi:2013.direct.collapse.smbh.bar.drainage.self.grav.disk.sims} explicitly simulated such bar drainage. Our own work on this mechanism \citep{hopkins:zoom.sims,hopkins:inflow.analytics,hopkins:qso.stellar.fb.together,hopkins:m31.disk,prieto:2016.zoomin.sims.to.fewpc.hydro.cosmo.highz,prieto:2017.zoomin.sims.agn.fueling.sne.fb,angles.alcazar:grav.torque.accretion.cosmo.sim.implications,daa:20.hyperrefinement.bh.growth} has argued that this is indeed the dominant fueling channel on $\sim 10\,{\rm pc}$--$10\,{\rm kpc}$ scales, and on scales where the stellar density dominates the potential.

However, as emphasized in those works themselves, what allows the bar/stellar-gravitational-torque mechanism to operate is that the driving stress is {\em external} to the disk gas -- it comes from stars (or dark matter) that dominate the local mass and density. This is by definition the regime $M_{\star}(<R) \gtrsim M_{\rm gas}(<R)$, i.e.\ where $\Delta_{\star} \gtrsim 1$. At the maser and BLR radii considered here ($R \sim 10^{-3}$--$1\,$pc), two independent observational facts rule this out as the dominant mechanism. First, the observed maximum stellar surface density in {\em any} known nuclear disk or nuclear star cluster (including the Galactic-center nuclear disk and the M31 nuclear disk, which are taken as canonical observed examples of the bar-drainage mechanism operating at the smallest resolvable scales) is $\Sigma_{\star} \lesssim 10^{5}\,{\rm M_{\odot}\,pc^{-2}} \sim 20\,{\rm g\,cm^{-2}}$ \citep{hopkins:maximum.surface.densities}. Extrapolating this bound inward using the \citet{hopkins:m31.disk} analytic nuclear-disk scalings, the stellar surface density that would be needed to supply $\Pi_{\rm ext}/\rho$ at the BLR radii is $\gtrsim 10^{8}\,{\rm M_{\odot}\,pc^{-2}}$, more than three orders of magnitude above the observed maximum. Second, and more decisively: the bar/stellar-torque mechanism {\em by definition} invokes stellar mass comparable to or exceeding the gas mass locally, which itself would produce a manifestly non-Keplerian rotation curve of exactly the form ruled out by the kinematic data compiled here. So the mechanism cannot be operating at radii where $V_{c}$ remains Keplerian to within $\lesssim 1-10\%$. We therefore agree with \citet{shlosman:inefficient.viscosities,begelman:direct.bh.collapse.w.turbulence,choi:2013.direct.collapse.smbh.bar.drainage.self.grav.disk.sims} and our own prior work that external-stress-driven drainage is an important mechanism at $\gtrsim$\,BHROI scales -- but at the maser and BLR radii interior to this, the mechanism cannot be dominant and the internal disk stress must supply the required $\Pi_{\rm eff}/\rho$.

\section{Conclusions}
\label{sec:conclusions}

Kinematics of masers and the broad-line region strongly constrain the allowed masses and mass profiles of the outer accretion disk around accreting supermassive black holes. For any self-consistent accretion disk model (where the stresses driving accretion cannot exceed the total stress/pressure in the disk), we show that this translates to constraints on the physics of what dominates the pressure in the accretion disk, and rule out many models immediately, regardless of free parameters in the models.

We specifically show that \textit{these constraints immediately rule out standard thermal-pressure dominated (``$\alpha$'') disks akin to those in SS73}. Any self-consistent thermal pressure-dominated disk would have a mass larger than the SMBH and a steeply-rising rotation curve, clearly ruled out by the data. Even if we arbitrarily fit the temperature and $\alpha$ parameter of the disk as a function of radius so as to provide any desired pressure profile, we show this would predict disk temperatures so hot that (1) maser and BLR emission would be impossible and the spectrum of the AGN would be completely incorrect, and (2) the thermal emission would more than ten orders of magnitude larger than observed. The predicted gas densities, optical depths, and many other properties are also inconsistent with the fact that we see maser and BLR emission at these radii. We similarly immediately rule out any disk which is dominated by radiation or cosmic ray pressure at these radii.

Magnetic-pressure dominated disks, specifically recently-proposed models of ``flux-frozen'' hyper-magnetized ($\beta \ll 1$) disks, on the other hand, are consistent with the present observations. The predicted gas densities and temperatures from such theoretical disk models are also in agreement with those needed for the BLR and maser emission.

It is plausible that turbulence could provide an order-unity fraction of the total pressure (unlike thermal or radiation pressure). However, models where turbulence ``alone'' provides the pressure (without e.g.\ strong magnetic fields) are strongly constrained: we can immediately rule out  models where such turbulence is powered by stellar feedback or by gravitational instabilities (``gravito-turbulent'' or ``marginally self-gravitating'' disks). While there is sufficient energy in gravity to power turbulence, in the absence of other appreciable pressures like magnetic pressure, we show that gravito-turbulent models would have to self-regulate at more like a turbulent Toomre $Q$ parameter of $\gtrsim 100$ (as opposed to the canonical $\sim 1$) in the outer disk, to be consistent with the observations.

We also show that, counter-intuitively, the self-consistent models for magnetically-dominated disks predict \textit{weaker} absolute magnetic field strengths than either (a) thermal-pressure dominated $\alpha$-disk models where the magnetic fields are provided by the MRI or the $\alpha$ arises from any Maxwell stress larger than or comparable to the Reynolds stress, or (b) gravito-turbulent disks with super-sonic and super-\Alf{ic} turbulence with Toomre $Q\sim 1$ driven by gravitational instabilities providing the Reynolds stress (assuming standard super-\Alf{ic} turbulent saturation). This is because even though the magnetic fields are \textit{relatively} more important in the magnetically-dominated disks, the \textit{absolute} pressures required in the disk to supply the observed accretion rates are orders-of-magnitude lower. The thermal-pressure-dominated disks, in particular, appear to also be independently ruled out by existing observations of maser Zeeman splitting. Surprisingly, \textit{stronger upper limits on magnetic fields favor more magnetically-dominated disks}.

Together, this appears to strongly favor the hypothesis that the \textit{outer} disks around AGN (at radii $R \gtrsim 1000\,R_{G} \sim 0.01\,$pc to the BH radius of influence at $\gtrsim $\,pc) are in a magnetically-dominated, flux-frozen state. Improved kinematic constraints can strengthen this conclusion and apply it to an even larger range of AGN. At much larger radii ($\gg$\,pc), while of course it is still interesting to understand what dominates the pressure and thermal structure of the ISM, it is not meaningful to speak of ``accretion disk'' solutions, since the gas is fully in the star-forming ISM and does not ``feel'' the BH potential and the dominant stresses can be extrinsic to the gas (from e.g.\ stellar bars or spiral arms or supernovae or colliding winds). 

At much smaller radii $\sim 1-100\,R_{G}$, where most of the thermal emission from the AGN originates, SS73-like models are most often assumed. While some of the BLR constraints we consider reach radii as small as $\sim 0.003\,$pc (a few light-days) and $R \sim 200-300\,R_{G}$ (around more massive BHs), unfortunately even much more precise measurements of rotation curves alone will not distinguish between models effectively at much smaller radii. This is because, although models here predict very different accretion disk masses even as $R \rightarrow 0$, they all predict the disk mass is very small compared to the BH mass ($M_{\rm disk} \ll M_{\rm BH}$) at such small radii, so gravitational deviations from Keplerian motion become tiny and basically undetectable. However, the different models also predict orders-of-magnitude different midplane densities, optical depths, accretion timescales, and scale heights $H$ for the disks: these could potentially be measured directly (rather than inferred indirectly through kinematics as we do here), with sufficiently high-resolution data, and would provide powerful model discriminants.

Two further observational tests are worth flagging. First, the inferred inflow timescale in hyper-magnetized disks, $t_{\rm acc} \sim 10\,\Omega^{-1}$, is orders of magnitude shorter than the canonical $\alpha$-disk viscous time $t_{\rm acc} \sim \alpha^{-1}\,(V_{c}/c_{s})^{2}\,\Omega^{-1}$. This is consistent with the growing observational evidence that ``changing-look'' and ``extreme-variability'' AGN undergo state transitions on timescales much shorter than predicted by standard $\alpha$-disks \citep{lamassa:2015.changing.look.agn,stern:2018.changing.look.agn.often.accretion.disk.state.changes,rumbaugh:2018.extreme.variability.qsos.often.lower.mdot.state.changes,noda:2018.changing.look.agn.from.mdot.changes.other.pressure.important,macleod:2019.changing.look.quasars.blr.turning.on.off,ricci:2023.agn.changing.look.variability.mini.review,dong:2024.changing.look.agn.preferentialy.at.percent.eddington.near.state.change,jana:2025.state.change.changing.look.agn.too.fast.for.standard.viscous.time.more.like.dynamical}; in particular \citet{jana:2025.state.change.changing.look.agn.too.fast.for.standard.viscous.time.more.like.dynamical} explicitly argue that the observed state-change timescales require a {\em dynamical}, not viscous, timescale, in natural agreement with the hyper-magnetized picture. Second, if the disk is fed episodically from the ISM at the BHROI, its angular momentum is set by whichever single GMC or gas parcel most recently accreted. In the ``superzoom'' simulations of \citet{hopkins:superzoom.overview,hopkins:superzoom.disk} the disk angular momentum is {\em not} tightly correlated with larger-scale galactic angular momentum and varies substantially episode-to-episode. So misalignment between the AGN disk and the host-galaxy stellar/gas kinematics is expected rather than problematic, but a systematic survey of BLR/maser-disk orientations vs. host-galaxy angular momenta would provide a valuable additional test.

One important question is whether the systems studied here (for which interesting kinematic constraints are available) actually form a representative subsample. For example, in principle one might argue that standard $\alpha$-disks, apparently ruled out here, are prevalent in other AGN at large radii but somehow cannot produce masers, so only magnetically-dominated disks would be represented in those samples. However this seems unlikely. First, we obtain qualitatively consistent constraints from a wide variety of methods studying different types of objects: maser kinematics, resolved interferometry of nearby (lower-luminosity) BLRs, microlensing, BLR reverberation mapping, and direct imaging of neutral gas disks. It seems implausible that all of these suffer from the same biases. Second, there is no evidence that any of the sub-populations for which these constraints are measured are highly biased from the general population of quasars in other properties that might be indicative of the type of accretion disk, like their thermal continuum spectral energy distributions. Still, this provides further motivation to expand all of these samples to more diverse AGN populations.

\begin{acknowledgements}
Support for PFH was provided by a Simons Investigator Grant.
\end{acknowledgements}

\bibliographystyle{mn2e}
%\bibliography{/Users/phopkins/Dropbox/Public/ms}
\bibliography{ms_extracted}

\end{document}